\documentclass[sigconf,9pt]{acmart}

\usepackage{soul}
\usepackage{url}
\usepackage[utf8]{inputenc}
\usepackage{graphicx}
\usepackage{amsmath}
\usepackage{booktabs}
\usepackage{subfig}
\usepackage{algorithm}
\usepackage{amsmath}

\usepackage{amssymb,amsfonts}
\usepackage{algorithmic}
\usepackage{graphicx}
\usepackage{textcomp}
\usepackage{float}
\usepackage{bm}
\usepackage{soul}
\usepackage{xcolor}
\usepackage{physics}
\usepackage{bbm}
\usepackage{lipsum}
\usepackage{multicol}
\usepackage{mathrsfs}
\usepackage{soul}

\makeatletter
\newcommand\thefontsize{The current font size is: \f@size pt}
\makeatother

\AtBeginDocument{%
	\providecommand\BibTeX{{%
			\normalfont B\kern-0.5em{\scshape i\kern-0.25em b}\kern-0.8em\TeX}}}

\setcopyright{acmcopyright}
\copyrightyear{2020}
\acmYear{2020}

\acmConference[SenSys '20]{The 18th ACM Conference on Embedded Networked Sensor Systems}{November 16--19, 2020}{Virtual Event, Japan}
\acmBooktitle{The 18th ACM Conference on Embedded Networked Sensor Systems (SenSys '20), November 16--19, 2020, Virtual Event, Japan}
\acmPrice{15.00}
\acmDOI{10.1145/3384419.3430735}
\acmISBN{978-1-4503-7590-0/20/11}

\begin{document}
	
	\title{RF-Net: A Unified Meta-Learning Framework for RF-enabled One-Shot Human Activity Recognition}
	\author{Shuya Ding, Zhe Chen, Tianyue Zheng, and Jun Luo}
	\authornote{The first two authors contribute equally to this research. Our open-source codes and datasets can be found at \url{https://github.com/di0002ya/RFNet.git}.}
	\affiliation{
	    School of Computer Science and Engineering, Nanyang Technological University, Singapore \\
	    Email: \{di0002ya, chen.zhe, tianyue002, junluo\}@ntu.edu.sg
	}
	


	
	\newcommand{\systemname}{RF-Net$ $}

	\begin{abstract}
		Radio-Frequency~(RF) based device-free Human Activity Recognition~(HAR) rises as a promising solution for many 
		applications.
		%
		However, device-free (or contactless) sensing is often more sensitive to environment changes than device-based (or wearable) sensing. Also, RF datasets strictly require on-line labeling during collection, starkly different from image and text data collections where human interpretations can be leveraged to perform off-line labeling. Therefore, existing solutions to
		RF-HAR entail a laborious data collection process 
		for adapting to new environments.
		To this end, we propose \systemname\ as a meta-learning based approach to one-shot RF-HAR; it reduces the labeling efforts for environment adaptation to the minimum level. In particular, we first examine three representative RF sensing techniques and two major meta-learning approaches. The results motivate us to innovate in two designs: i) a dual-path base HAR network, where both time and frequency domains are dedicated to learning powerful RF features including spatial and attention-based temporal ones, and ii) a metric-based meta-learning framework to enhance the fast adaption capability of the base network, including an RF-specific metric module along with a residual classification module. We conduct extensive experiments based on all three RF sensing techniques in multiple real-world indoor environments; all results strongly demonstrate the efficacy of \systemname\ compared with state-of-the-art baselines.
	\end{abstract}

	
	
	
	\maketitle

	\section{Introduction} \label{sec:intro}
	Recently, Human Activity Recognition (HAR) has been attracting attention increasingly in both \textcolor{black}{academia} and industry, giving its promising potential in real-world applications such as smart home~\cite{b1_smart_home,b2_smart_home,chi2018ear}, health care~\cite{b3_health_care,keally2011pbn,ghose2013unobtrusive}, and fall detection~\cite{b6_CVPR_detection,wang2014wifall,b5_fall_detection}. Generally, two categories for HAR have been explored: \textit{device-based} (or wearable) and \textit{device-free} (or contactless). Device-based HAR leverages the wearable devices such as smart phones or watches to recognize human activities~\cite{keally2011pbn,truong2018capband,b7_qian,yuanhe2017pervasive}. However, it may cause discomfort and extra burden, which leads to the alternative method of device-free HAR. This later method exploits camera-based image data~\cite{hu2015jointly,ma2016going,liu2019caesar}, acoustic/ultrasonic signals~\cite{hao2013isleep,BreathListener-MobiSys19,wang2019millisonic}, and Radio-Frequency~(RF) signals~\cite{chi2018ear,b15_csi,b32_widar3} to achieve HAR.
	%
	\begin{figure}[t]
		\vspace{-1.5ex}
		\centering
		\subfloat[Wiping (time)]{
		    \begin{minipage}[b]{0.48\linewidth}
		        \centering
			    \includegraphics[width = .96\textwidth]{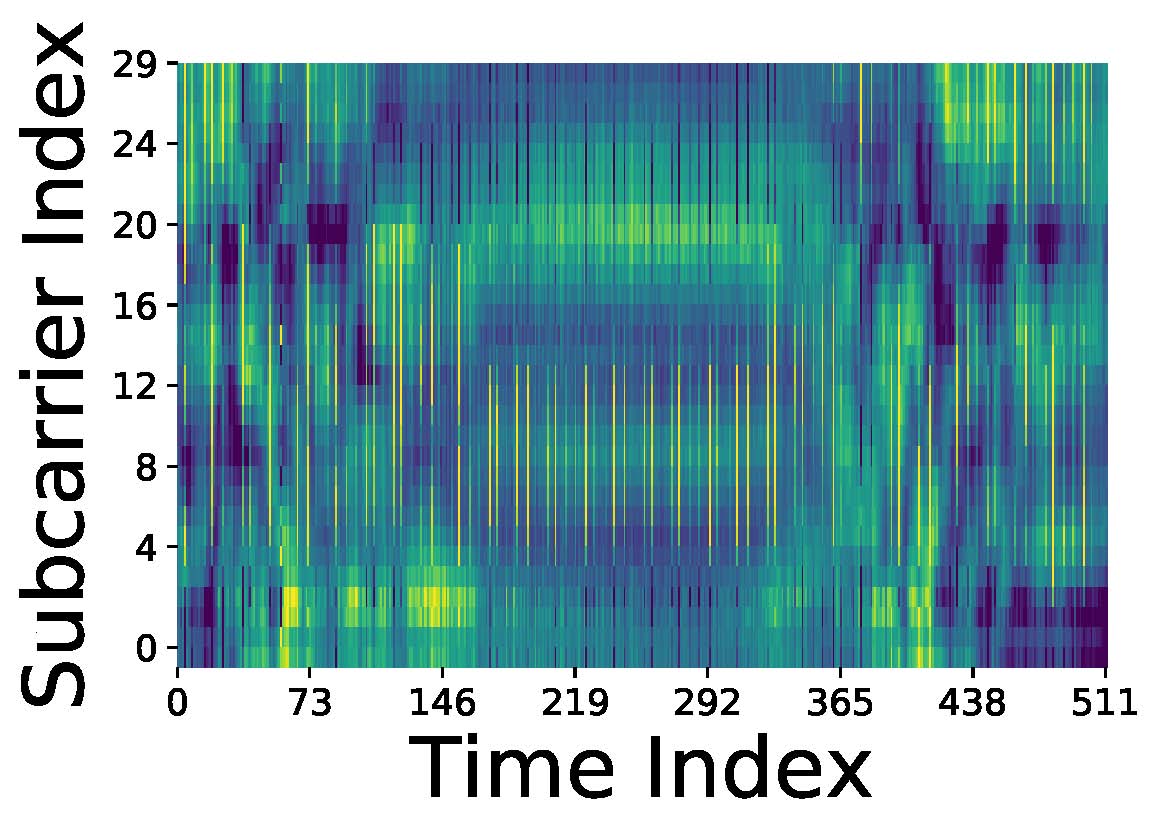}
			    \vspace{-.5ex}
			\end{minipage}
        }%
		\subfloat[Walking (time)]{
		    \begin{minipage}[b]{0.48\linewidth}
		        \centering
			    \includegraphics[width = .96\textwidth]{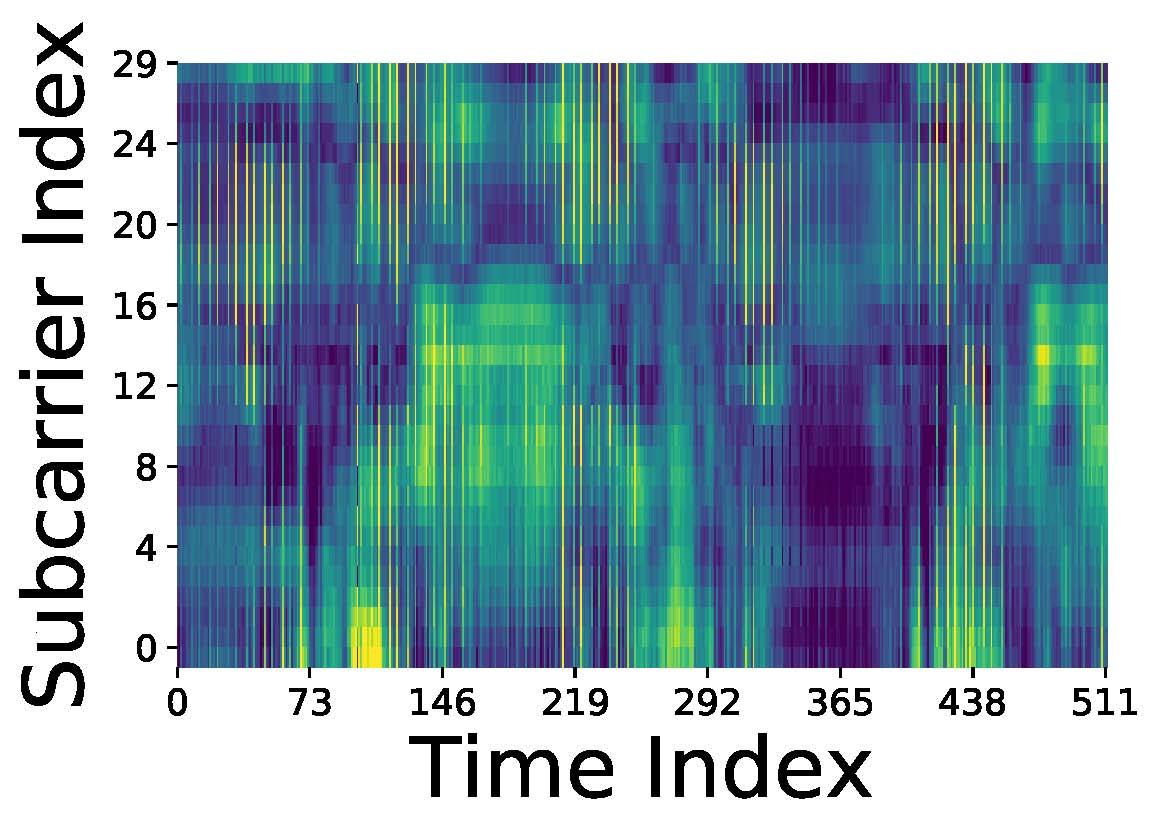}
			    \vspace{-.5ex}
			\end{minipage}
		}\\\vspace{-2ex}
		\subfloat[Wiping (frequency)]{
		    \begin{minipage}[b]{0.48\linewidth}
		        \centering
			    \includegraphics[width = .96\textwidth]{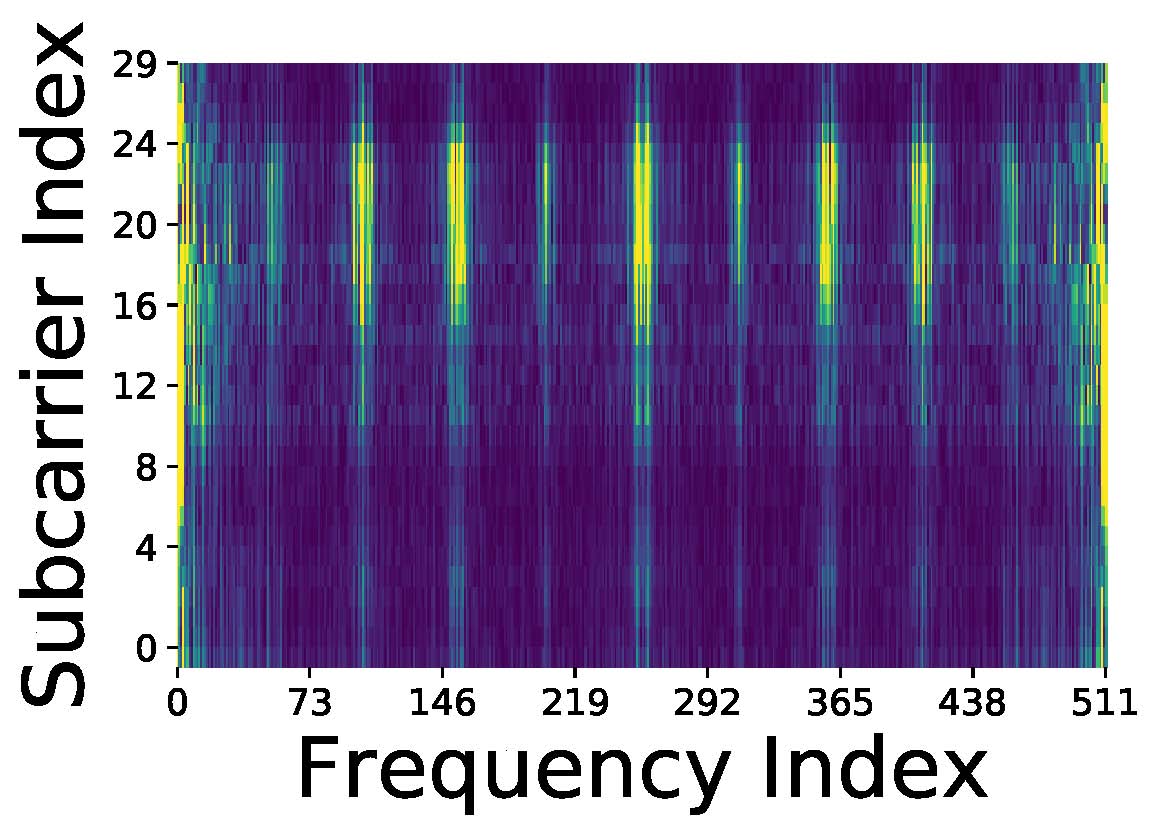}
			    \vspace{-.5ex}
			\end{minipage}
		}%
		\subfloat[Walking (frequency)]{
		    \begin{minipage}[b]{0.48\linewidth}
		        \centering
			    \includegraphics[width = .96\textwidth]{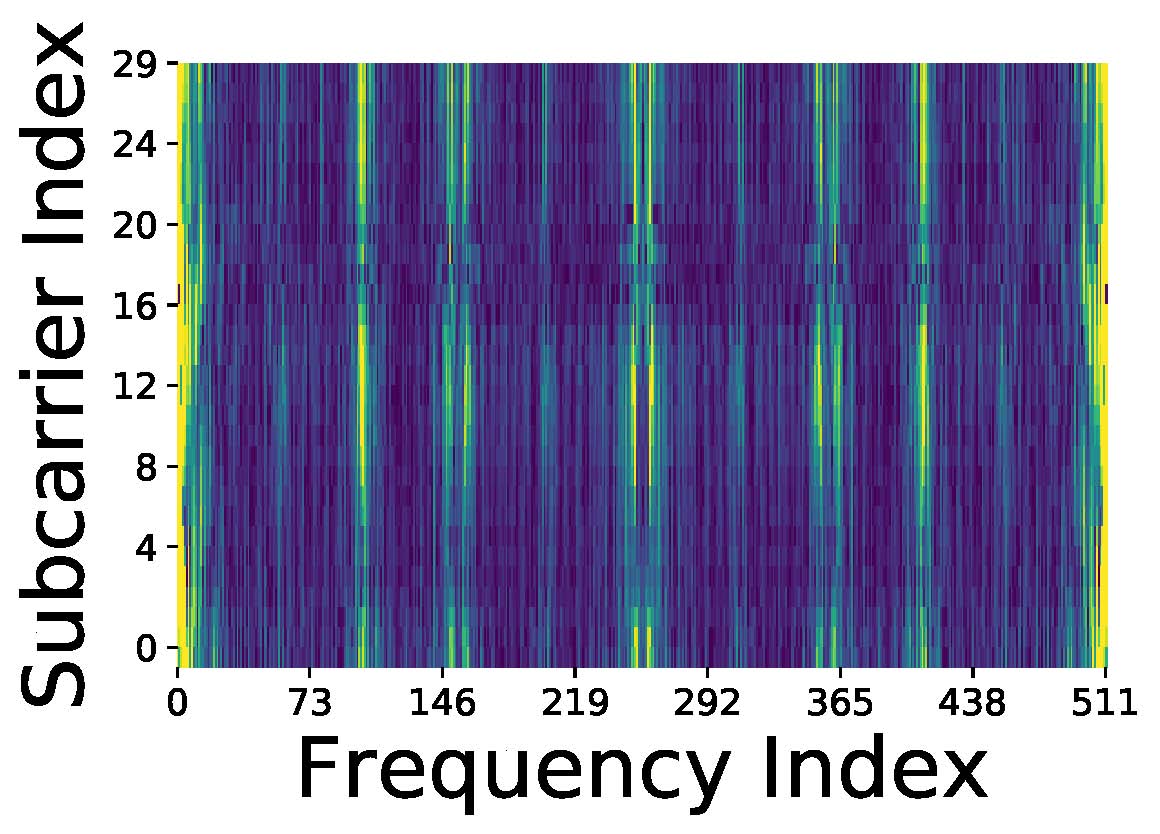}
			    \vspace{-.5ex}
			\end{minipage}
		}
	    \vspace{-1ex}
		\caption{Wi-Fi CSI heatmaps of two different activities: wiping the board and walking around. It is virtually impossible to intuitively recognize them by human eyes.}
		\label{fig:sensitive}
        \vspace{-2ex}
	\end{figure}
	
	Research efforts leveraging the efficacy of deep learning have been intensively carried out on camera-based approaches~\cite{b9_Lin_2019_ICCV,liu2019caesar,ma2016going}, but exploiting image data bears a risk of privacy infringement especially in private places~\cite{b1_smart_home}. 
	Meanwhile, acoustic/ultrasonic sensing often requires bulky devices and incurs a high energy consumption, hence rarely adopted in practice even though the performance can be excellent in noise-free environments~\cite{hao2013isleep,BreathListener-MobiSys19}.
	Fortunately, RF sensing offers just enough resolution to perform HAR without infringing personal privacy, and it is far more energy-efficient than sound, as RF signals are not generated by mechanical vibration.
	Moreover, its performance is robust even under various temperature or lighting conditions. Therefore, RF-HAR is deemed as the most promising solution, where Wi-Fi
	is often adopted.

	Early RF-HAR exploited RSSI (Received Signal Strength Indicator) to analyze human activities~\cite{b11_rss1,b12_rss2}, collected by COTS (Commercial-Off-The-Shelf) Wi-Fi cards. However, RSSI can only represent coarse information of Wi-Fi signals, rather than fine-grained \textcolor{black}{multipath} effects generated by human activities. Therefore, many efforts have been recently devoted to extracting comprehensive information in Wi-Fi CSI (Channel State Information) from Intel 5300~\cite{b10_intel5300} Wi-Fi cards~\cite{b13_wei_wang,b14_csi,b15_csi,b16_csi}. 
	As illustrated in Figure~\ref{fig:sensitive}, body movements of different activities bring variations to CSI, so features generated from CSI could be leveraged for HAR. Whereas Wi-Fi sensing exploits devices originally designed for wireless communication purposes (hence inherently limited in sensing performance), dedicated RF sensing techniques are gaining momentum very recently. These techniques are typically supported by two types of radios, namely FMCW (Frequency-Modulated Continuous Wave) radio~\cite{b35_3d,b28_witrack} and \textcolor{black}{impulse radio~\cite{kim2017hand,park2016ir,v2ifi}.}
	
	Although RF sensing, in general, has achieved great performance in HAR, it does have two major weaknesses. On one hand, similar to other device-free sensing techniques~\cite{BreathListener-MobiSys19,mcintosh2017echoflex,liu2019caesar}, RF sensing requires high volume of training data to re-train its model if the sensing environment is altered (e.g., changing the radio locations and/or the room furniture layout). This weakness is inherent to device-free sensing as, unlike device-based sensing, it senses both the subject and the background. On the other hand, as illustrated in Figure~\ref{fig:sensitive}, human cannot intuitively recognize different activities from RF sensing data (as opposed to images). Consequently, human interpretation cannot be exploited for conducting off-line labeling, which is essential for image and text data collections. As a result, RF sensing has to endure a much more laborious process (compared with computer vision and natural language processing) to gather labeled data sufficiently, in order for RF-HAR approaches to be able to adapt to new environments with satisfactory performance.

	Fortunately, recent developments on meta-learning~\cite{santoro2016meta,b26_maml,b18_matching_net} have offered us a chance to enhance the environment adaptivity of RF sensing. Roughly categorized into three types: model-based, optimization-based, and metric-based, meta-learning aims to \textit{adapt to new tasks rapidly with few labeled observations}. 
	Whereas model-based approaches~\cite{santoro2016meta,munkhdalai2017meta} 
	incur a high computational complexity, the other two may potentially help RF-HAR.
	Optimization-based approaches~\cite{b26_maml,meta_leo,meta_taml} adjust the training algorithm to find good initialization weights or learning rate, but their generalization ability is questionable and the incurred overhead can still be high~\cite{sun2019meta,meta_leo}.
	Metric-based approaches~\cite{b23_siamese,b18_matching_net,b19_proto_net} classify an unlabeled observation by its similarity to the labeled data. 
	As they focus on learning an optimal similarity metric 
	(instead of directly tuning the learning architecture), they likely incur the lowest complexity.

	

\begin{figure}[t]
    \centering
    \includegraphics[width=0.48\textwidth]{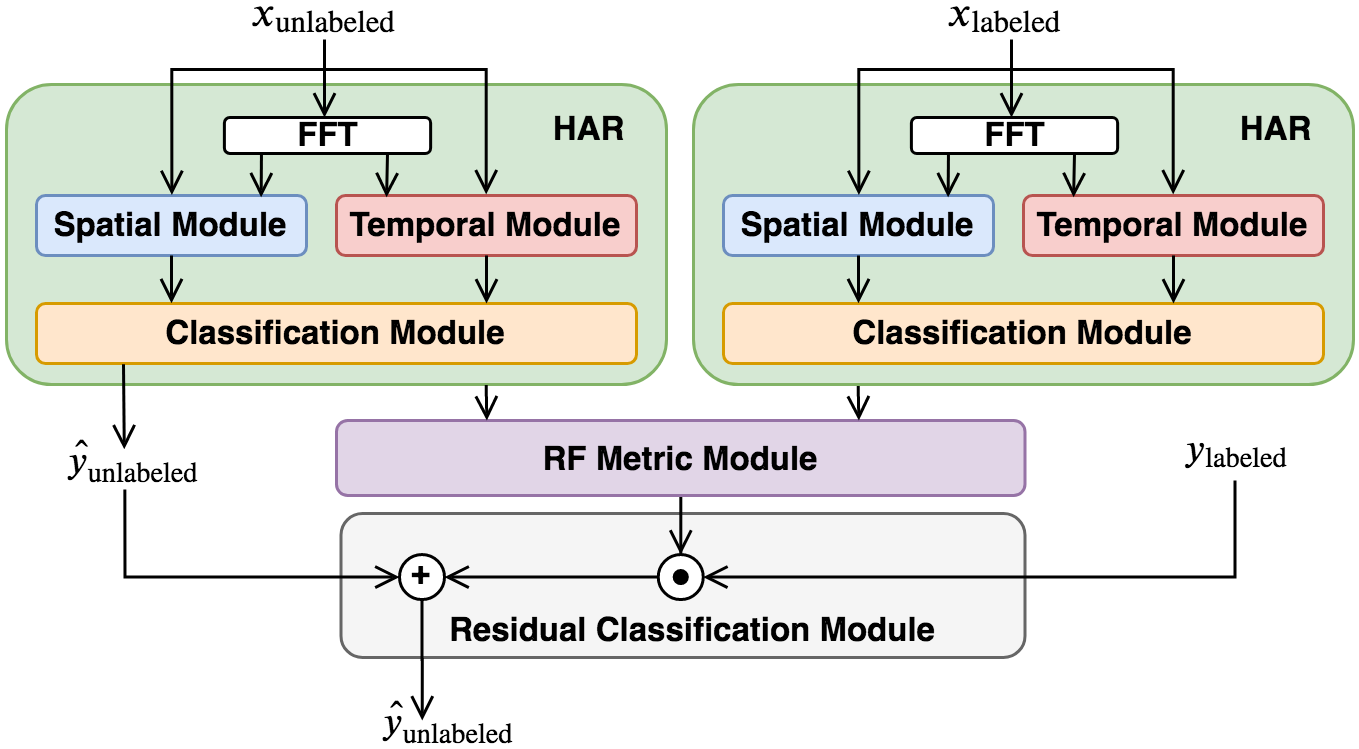}
    \vspace{-4ex}
    \caption{{\systemname} overview.}
    \label{fig:simp_overview}
\end{figure}

	Leveraging the power of meta learning, we propose {\systemname} \textcolor{black}{for} one-shot RF-HAR. In other words, {\systemname} performs HAR accurately in a new environment with only one observation for each label. Specifically, we first carefully examine representative RF sensing techniques along with major meta-learning approaches. The results motivate us to innovate in two aspects: i) a dual-path base network for classifying activities, and ii) a metric-based meta-learning framework to improve the fast adaption capability of the base network, as illutrated in Figure~\ref{fig:simp_overview}. For the base HAR network, we combine spatial module along with attention-based temporal module, aiming to learn influential spatial-temporal features from both time and frequency domains. Our meta-learning framework contains a parametric RF-specific module designed to train a powerful distance metric,
	instead of simply relying on traditional non-parametric metrics (e.g., Euclidean or Cosine distances). This helps to achieve a better generalization when applying this metric to 
	conduct classifications in new environments.
	Essentially, we employ the base network to perform both activity recognition and feature extraction, and exploit them to (meta)-train the distance metric via a residual classification module. 
    %
	%
	In summary, our major contributions are: 
	\begin{itemize}
		\item We propose {\systemname}, a unified meta-learning framework for RF-enabled one-shot HAR, delivering the capability of being adaptive to new environments with very few labeled data.
		\item We innovatively design a dual-path activity recognition (base) network, aiming to learn influential features from general RF signals for enhancing HAR accuracy.
		\item We equip our meta-learning framework with a novel RF-specific module designed to train a powerful distance metric, so as to achieve a better generalization.
		\item We conduct extensive experiments on multiple RF sensing techniques and in many indoor environments. We demonstrate the superior performance of our proposed {\systemname} compared against multiple baselines.
	\end{itemize}
	This paper is organized as follows. Background and related works are first examined in Section~\ref{sec:bg}. Then our \systemname\ is presented in Section~\ref{sec:metaRF}. Extensive experiment results are reported in Section~\ref{sec:eval}, and our paper is concluded in Section~\ref{sec:con}.

	\section{Background and Literature}\label{sec:bg}
	We carefully study representative RF sensing techniques along with major meta-learning approaches in this section, aiming to better motivate our design. The rationale behind this study is that, as RF sensing induces far more complicated input data than those under wearable sensing (where only up to three time series from the three axes of an IMU sensors present~\cite{kempe2011inertial}), deep understanding and innovative treatments are hence needed.

	\begin{figure*}[t]%
		\centering
		\subfloat[RF signal matrix.] 
		{\label{fig:sig_mat} 
		    \begin{minipage}[b]{0.24\linewidth}
		        \centering
			    \includegraphics[width = .98\textwidth]{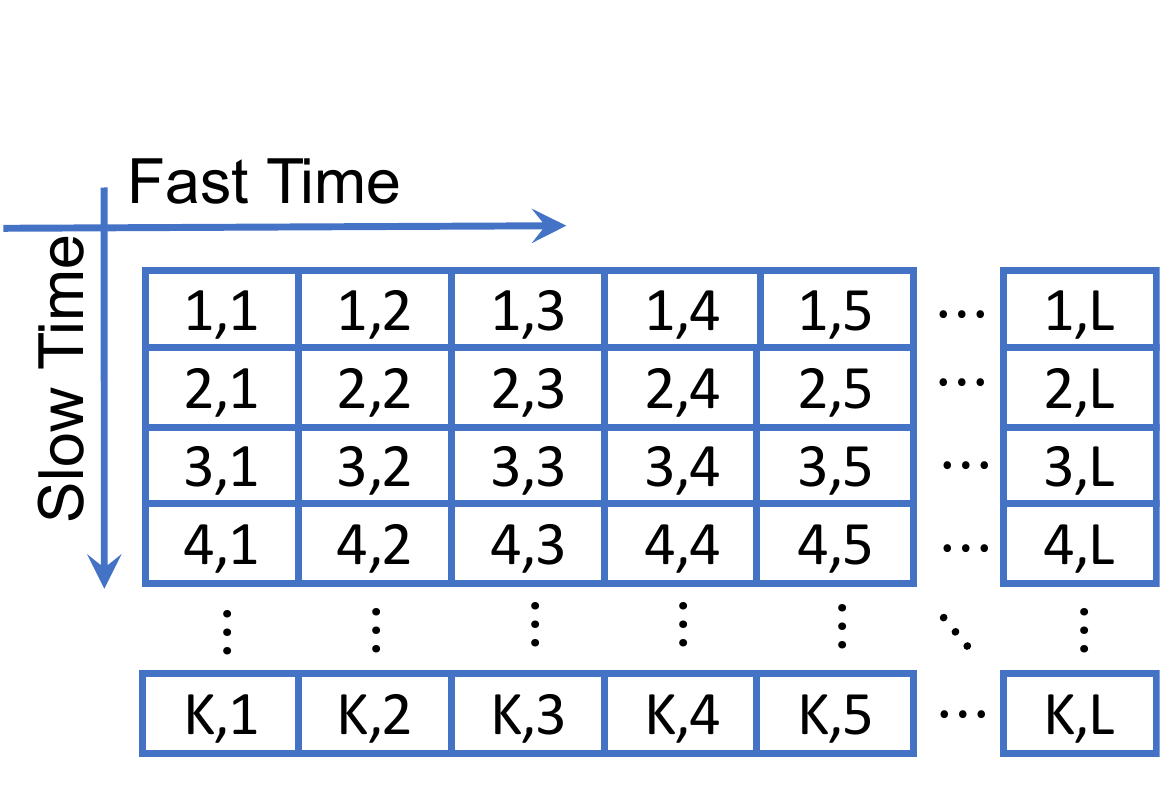}
			\end{minipage}
		}
		\hfill
		\subfloat[Wi-Fi matrix.]
		{\label{fig:mat_wifi}
		    \begin{minipage}[b]{0.24\linewidth}
		        \centering
			    \includegraphics[width = .86\textwidth]{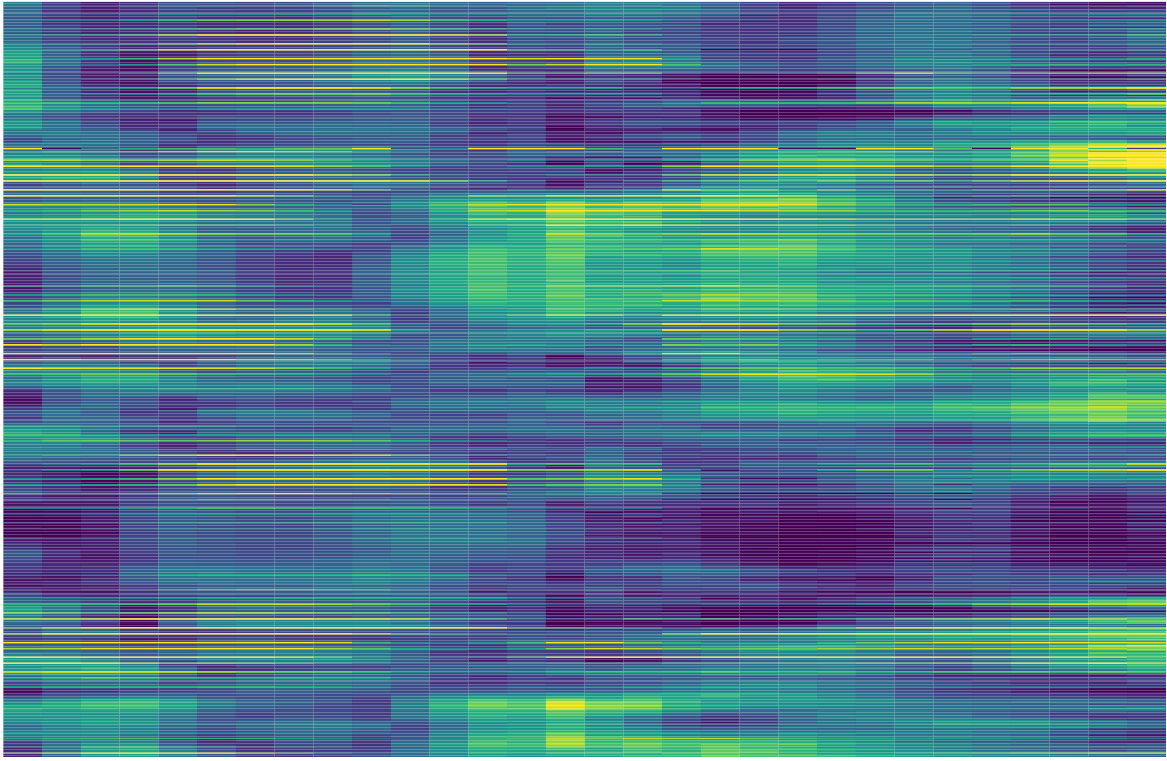}
			\end{minipage}
		}
		\hfill
		\subfloat[FMCW matrix.]{
			\label{fig:mat_fmcw}
			\begin{minipage}[b]{0.24\linewidth}
		        \centering
			    \includegraphics[width = .86\textwidth]{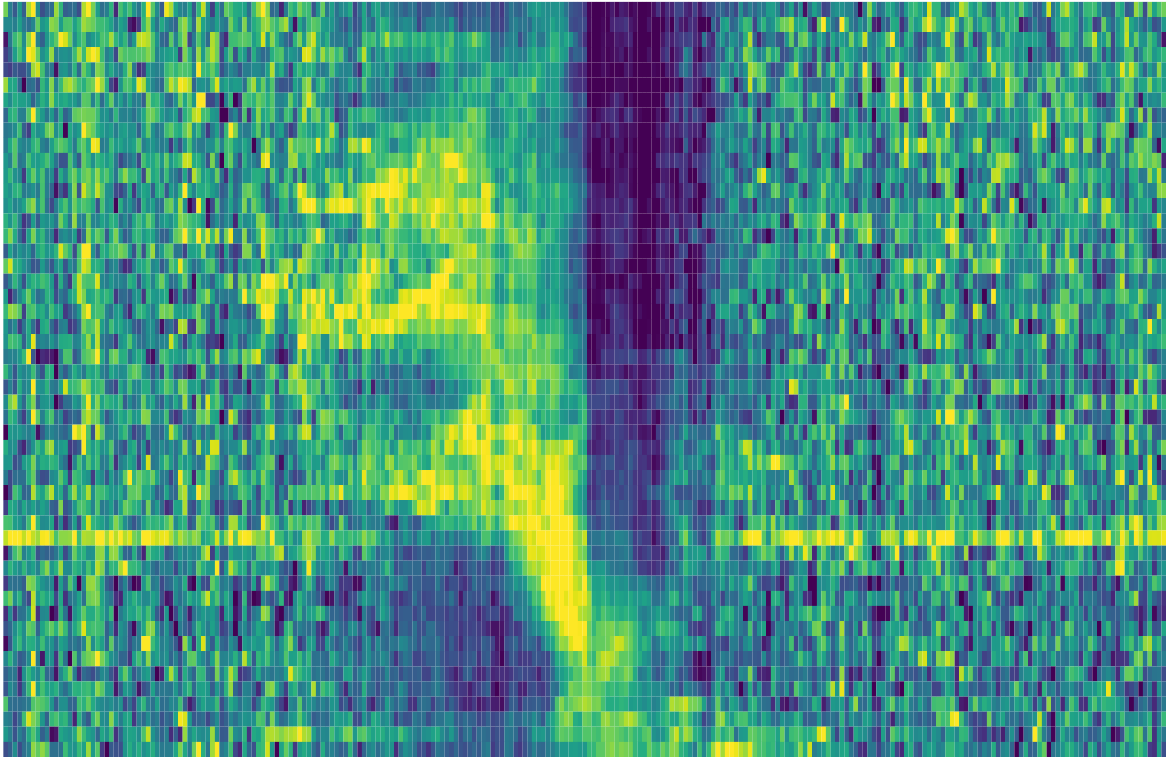}
			\end{minipage}
		}
		\subfloat[IR matrix.]
		{ \label{fig:mat_uwb}
		    \begin{minipage}[b]{0.24\linewidth}
		        \centering
			    \includegraphics[width = .86\textwidth]{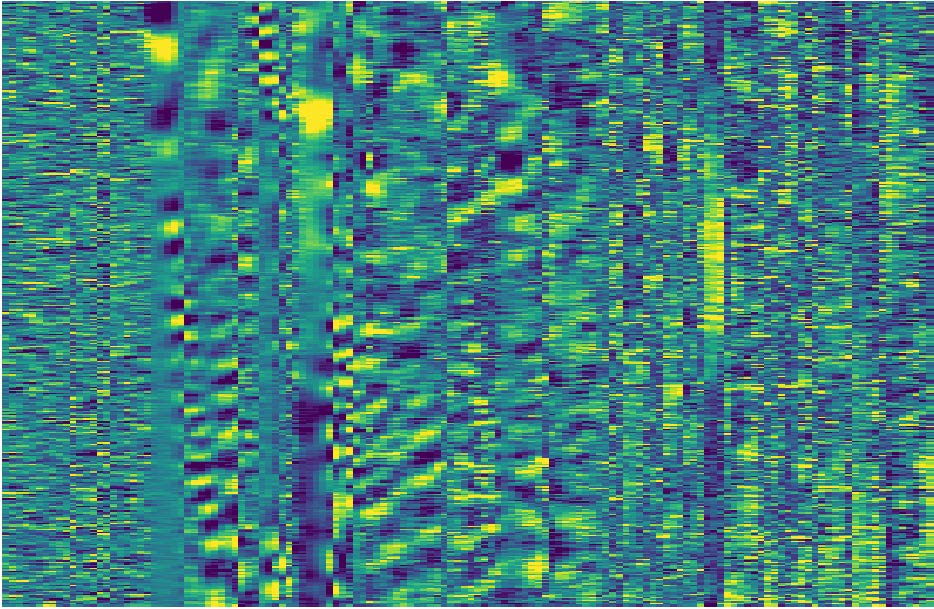}
			\end{minipage}
		}
	    \vspace{-1ex}
		\caption{The RF signal matrix and corresponding heatmaps for three RF techniques in terms of the “walking” activity. }
		\label{fig:heatmap}
		\vspace{-1ex}
	\end{figure*}
	
	\subsection{Modeling RF Sensing}\label{ssec:model}
	Generally, all RF sensing approaches explore RF \textit{channel state information} (CSI) to differentiate human activities. Therefore, we model the RF channel first, then examine three typical RF signals. According to~\cite{b30_tse}, considering a pair of transmitter and receiver in an indoor environment with $P$ propagation paths, we have the following baseband RF channel model given a carrier frequency $f_\mathrm{c}$:
	\begin{align}
		h(t) =  {\textstyle \sum_{p=1}^{P}} \alpha_{p} e^{j 2 \pi f_\mathrm{c} \tau_p }  + n(t),
	\end{align}
	where  $\alpha_{p}$ is the amplitude of $p$-th path, and $n(t)$ is Gaussian noise.  Moreover, $\tau_p = \tau_p^{S} + \tau_p^{D}$ where $\tau_p^S$ and $\tau_p^D$ are the $p$-th time delays caused by static reflections and motion reflections, respectively. For a transmitted signal $s(t)$, the received signal becomes $y(t) = (h * s)(t)$, where $*$ denotes convolution. In the following, we stick to only one path (thus removing the subscript $p$), and we omit noise term for brevity.

	\subsubsection*{Wi-Fi Radio}
	Wi-Fi communications utilize Orthogonal Frequency Division Multiplexing (OFDM) to encoding digital data on multiple subcarriers. 
	Let $L$ denote the number of OFDM subcarriers, $s_\ell$ denotes the $\ell$-th subcarrier, then the received signal $y_\ell^W$ can be represented as follows:
	\begin{align}
	y_\ell^\mathrm{W} =  \alpha_\ell e^{j 2 \pi f_\ell \tau } s_\ell  , \quad \ell \in \{  1,..., L \},
	\end{align}
	where $\alpha_\ell$ and $f_\ell$ are the amplitude and frequency of the $\ell$-th subcarrier, respectively. Here the time variable $t$ disappears because the bandwidth of Wi-Fi is so narrow that it is approximately deemed as time invariant. 
	The channel state can then be estimated as $\hat{h}_\ell = y_\ell^\mathrm{W} / s_\ell$, whose phase $\angle\hat{h}_\ell$ contains temporal feature $\tau$.
	%
	%
	%
	%
	Given a total $K$ received packets, each of them offers a CSI vector $[\hat{h}_\ell]_{\ell \in \{  1, \cdots, L \}}$. Combining all these data, we obtain the ensemble input sample as a $K \times L$ signal matrix. Here we term the row index as \textit{slow-time} and the column index as \textit{fast-time}, as they represent sampling at different temporal scales. Figure~\ref{fig:sig_mat} illustrates this signal model, while Figure~\ref{fig:mat_wifi} provides an example for a Wi-Fi matrix. 
	
	\subsubsection*{FMCW Radio}
	FMCW is a special type of radar implementation. Different from normal radios, FMCW radio transmits analog signals modulated in a continuously increasing frequency across a wide bandwidth.
	Consequently, it sweeps across time with a fine-grained (frequency) resolution, equivalently convertible to a fine-grained range resolution.
	For FMCW radio, we denote the bandwidth by $B$ and sweeping time span by $T^\mathrm{S}$. 
	According to~\cite{b30_radar_book}, the received FMCW signal $y^\mathrm{F}(t)$ is:
	\begin{align} \label{eq:rx_bb}
	y^\mathrm{F}(t)  =  \alpha \Pi (t-\tau) e^{-j2\pi\left( \beta \tau t + f_\mathrm{c} \tau - 0.5\beta \tau^2  \right) },
	\end{align}
	where $\beta = B/T^\mathrm{S}$ and $\Pi(t)$ is a rectangle function ranged from $-T^\mathrm{S}/2$ to $T^\mathrm{S}/2$. We term a signal represented by Eq~\eqref{eq:rx_bb} a \textit{frame}; it uses a frequency $\beta\tau$ and a phase $f_\mathrm{c}\tau-\beta\tau^2/2$ to represent how the signal changes over time. Therefore, we perform an $L$-point FFT on each frame and get a set of $L$ frequency components, which correspond to fast-time samples. Then we combine all $K$ frames together to construct a $K \times L$ signal matrix (similar to that of Wi-Fi), as shown in Figure~\ref{fig:mat_fmcw}. 
	
	\subsubsection*{Impulse Radio}
	Whereas FMCW radio uses a varying frequency to sweep time, impulse radio (IR) transmits a pulse signal with an extremely short time duration. This enables IR signals to also occupy a wide bandwidth $B$, again leading to a fine-grained time resolution.
	The received pulse signal $y^\mathrm{I}(t)$ is:
	\begin{align} \label{eq:tx_bb}
	y^\mathrm{I}(t) = \alpha  e^{j 2 \pi f_\mathrm{c} \tau} e^{- 0.5 \left(t - 0.5T_\mathrm{tx} - \tau \right)^2 \epsilon^{-2}_\mathrm{tx} },
	\end{align}
	where $T_\mathrm{tx} = \frac{1}{B}$ is the signal duration, and $\epsilon_\mathrm{tx} = \frac{ 1 }{ 2 \pi B_\mathrm{-10~\!dB} (\log_{10}(e))^{1/2}}$ is the standard deviation determining the -10 dB bandwidth. 
	Similar to FMCW, if we consider $L$ samples of a pulse as a \textit{frame}, combining $K$ such frames again gives us a $K \times L$ signal matrix, where the fast-time directly corresponds to the temporal sample indices of a frame, as shown in Figure~\ref{fig:mat_uwb}. 
	
	\vspace{1ex}
	\noindent\textbf{Remark}: If we deem this matrix as an image (common input to deep learning models such as CNN~\cite{lecun1998gradient,krizhevsky2012imagenet}), we may stack multiple such matrices derived from different tx-rx pairs together, so that each tx-rx pair corresponds to an input channel. Although these three RF sensing techniques share a similar model, they offer very different time (thus range) resolutions: the 20~\!MHz bandwidth of Wi-Fi can only translate to a range resolution about 15~\!m, but both FMCW and impluse radios, with more than 1~\!GHz bandwidth, can achieve a centimeter-level resolution instead.


	\subsection{RF Meets Learning: Status and Challenges}
	Existing RF-sensing solutions mostly leverage machine learning techniques to extract features and classify activities~\cite{b13_wei_wang, b14_csi,ma2018signfi,weiwang2015keystroke,Venkatnarayan2018mulituser,virmani2017position,b29_fmcw,b35_3d}.
	Although their adopted RF signals and designed algorithms vary, their input data can all be unified under the RF signal matrix introduced in Section~\ref{ssec:model} as subcategories.
	For example, the proposals in~\cite{b13_wei_wang, b14_csi,youssef2019ubiquitous,weiwang2015keystroke,Venkatnarayan2018mulituser,virmani2017position} extract Doppler shift only along the slow time axis in Figure~\ref{fig:mat_wifi}, in order to capture the velocities of moving parts of a target and in turn to perform classifications. Other proposals~\cite{b29_fmcw,b35_3d} utilize multiple tx-rx pairs to extract range information embedded along the fast time dimension in Figure~\ref{fig:mat_fmcw}, so as to predict human poses. Essentially, existing solutions only retrieve partial information provided by RF signals; this motivates us to consider a full exploitation of spatial, temporal, frequency features offered by the signal matrix, so as to handle all three RF sensing techniques using a unified learning framework.

	\begin{figure}[t]
	\vspace{-1ex}
	\centering
	\subfloat[Wi-Fi CSI in environment 1.]
	{
	    \begin{minipage}[b]{0.48\linewidth}
		\centering
		\includegraphics[width=.88\textwidth]{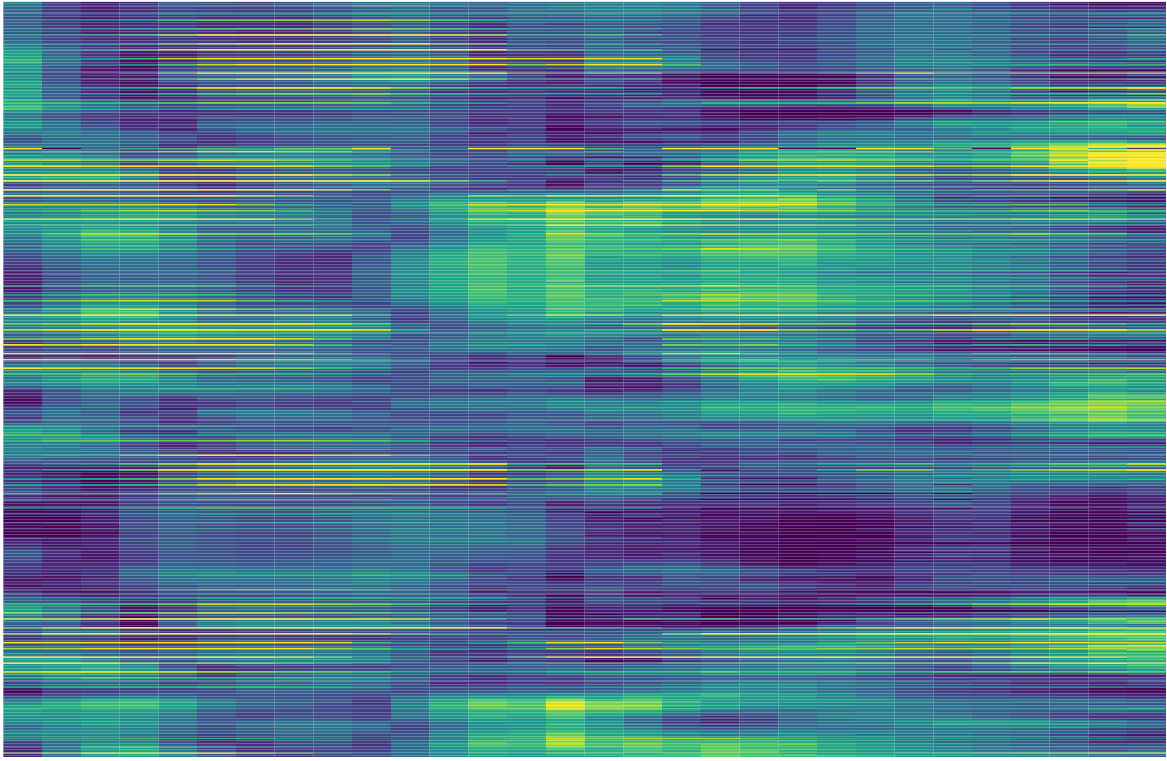}
		\label{fig:wifi_walk1}
		\end{minipage}
	}
	\subfloat[Wi-Fi CSI in environment 2.]
	{
	    \begin{minipage}[b]{0.48\linewidth}
		\centering
		\includegraphics[width=.88\textwidth]{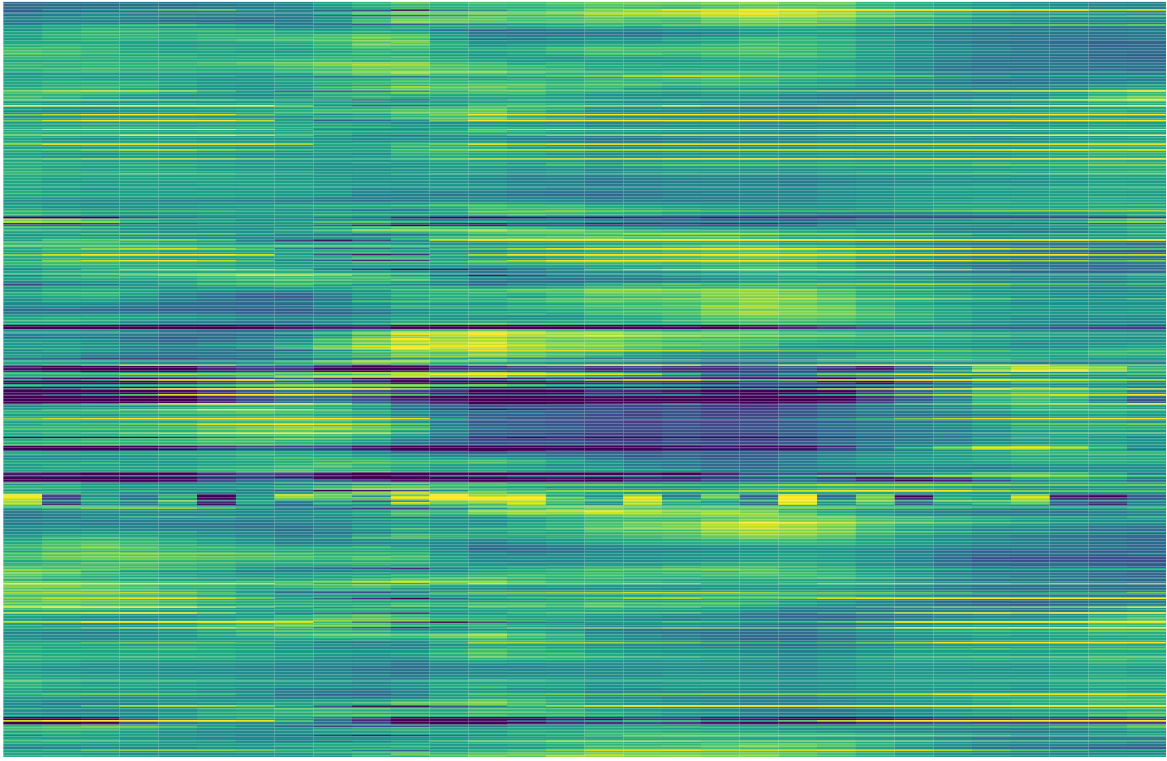}
		\label{fig:wifi_walk2}
		\end{minipage}
	}
	\\
	\vspace{-1ex}
	\subfloat[Feature map of environment 1.]
	{
	    \begin{minipage}[b]{0.48\linewidth}
		\centering
		\includegraphics[width=.88\textwidth]{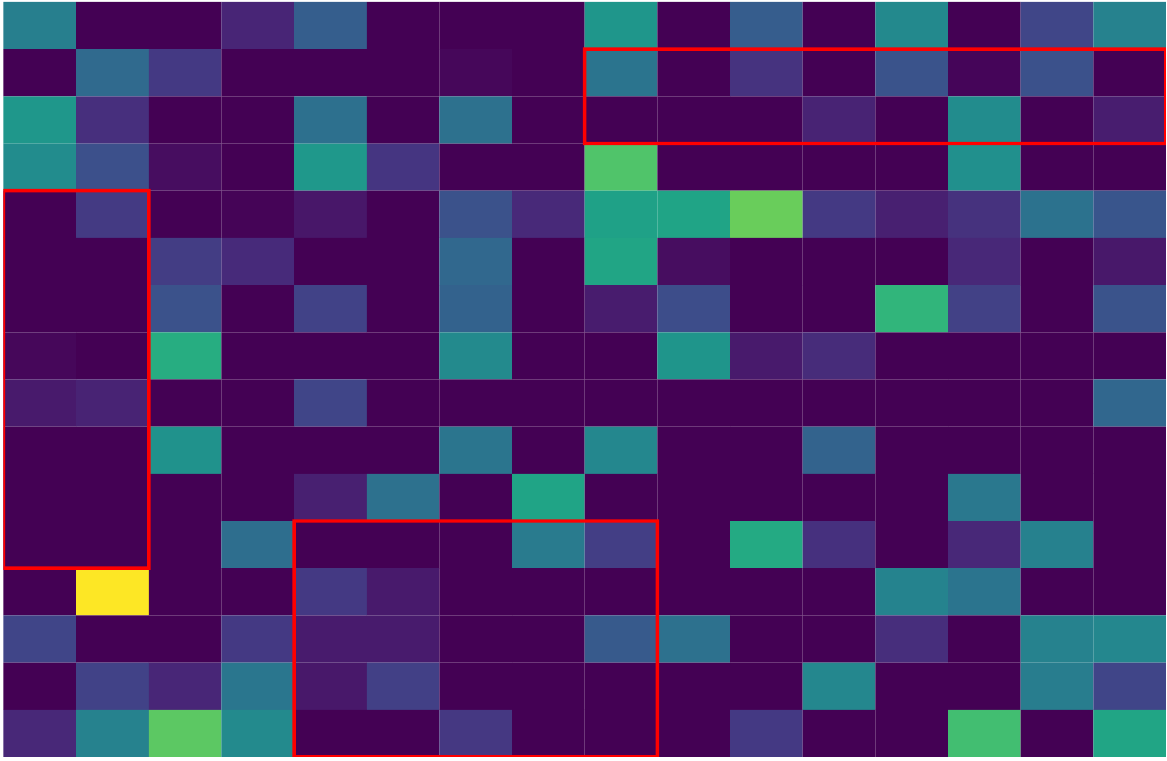}
		\label{fig:feature_map1}
		\end{minipage}
	}
	\subfloat[Feature map of environment 2.]
	{
	    \begin{minipage}[b]{0.48\linewidth}
		\centering
		\includegraphics[width=.88\textwidth]{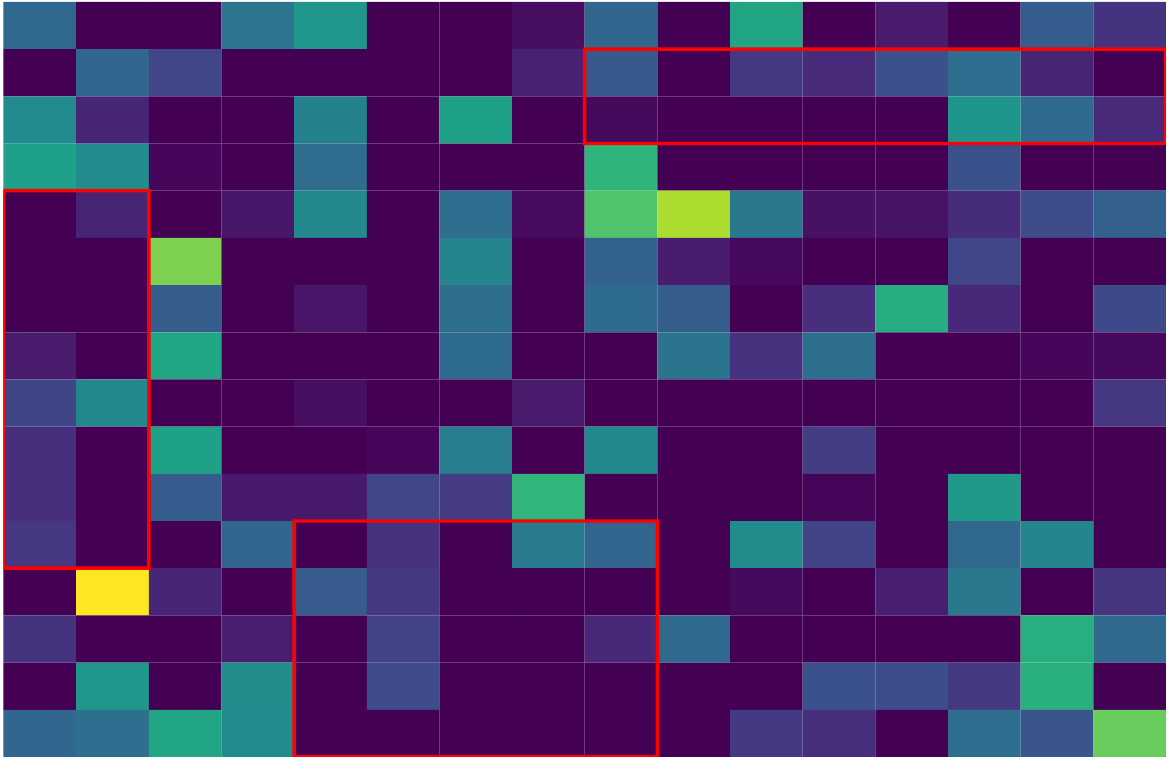}
		\label{fig:feature_map2}
		\end{minipage}
	}

	\vspace{-1ex}
	\caption{Environment influence: the same activity ``walking'' leads to distinct signal matrices and extracted features.}
	\label{fig:shape_comparison}
	\vspace{-1ex}
    \end{figure}

	
	However, exploiting more features is a double-edged sword: it improves activity classification under noise-free circumstances, but it also makes RF sensing more sensitive to environment changes in practice. Figure~\ref{fig:shape_comparison} \textcolor{black}{demonstrates} the influence of this sensitivity. As explained in Section~\ref{sec:intro}, re-training a learning model to suit a new environment can be extremely expensive, as RF signal matrices, totally different from images and texts, are not human understandable, enforcing a laborious on-line labelling. 
	One solution is to employ environment-invariant features/models, hoping to retain classification accuracy without re-training in a new environment. 
	Previous works~~\cite{b32_widar3, zhou2020towards, virmani2017position} utilize handcrafted features for this purpose. However, these features depend on prior information such as the position and orientation of a target, likely incurring another laborious process in obtaining these information.
	%
    %
    Another solution is transfer learning~\cite{b33_crosssense, b15_csi}. The basic idea is to learn ``transferable'' knowledge applicable across a pair of source and target environments. However, when applied to a new target environment, the whole fine-tuning process has to be performed again with a substantial amount of labeled data. In order to better tackle this challenge, we need to consider recently emerged alternatives.

	
	

	\subsection{Meta-Learning Basics}
	\label{ssec:metalearning}
	Human level intelligence requires that learning models can mimic human behavior to learn from known tasks (environments in our context) and adapt to new tasks quickly with only a few labeled observations. Recently, meta-learning~\cite{santoro2016meta,b26_maml,b18_matching_net} has emerged to achieve the aforementioned intelligence. The key idea of meta-learning is twofold: i) learning knowledge from source environments with rather abundant data, and ii) exploiting accumulated knowledge to learn similarities and differences in all target environments, requiring only a minimum level of labeled data. In meta-learning, \textcolor{black}{we denote the dataset for the $\ell$-th environment by $\mathcal{D}_\ell$}. Each $\mathcal{D}_\ell$ is split into a \textit{support set} $\mathcal{D}_\ell^\mathrm{S}$ for learning and a \textit{query set} $\mathcal{D}_\ell^\mathrm{Q}$ for training and testing. We drop the subscript $\ell$ in the following discussions as the learning procedure is identical in all source datasets. Generally, a \textit{base network} $f_\Phi$ predicts the probability $P_\Theta(y|\mathbf{x}, \mathcal{D}^\mathrm{S})$ of class $y$ in source environments, given a support set $\mathcal{D}^\mathrm{S}$ and an input $\mathbf{x}$ in $\mathcal{D}^\mathrm{S}$, where $\Theta$ is the \textit{meta-parameter} that parameterizes the conditional probability.
	\begin{align}
	\Phi^* = \arg\max_{\Phi}\mathbb{E}_{\mathcal{D}^\mathrm{S}} \left[ \sum \limits_{(\mathbf{x},y)\in \mathcal{D}^\mathrm{S}} P_{\Theta}(y|\mathbf{x},\mathcal{D}^\mathrm{S})\right].
	\end{align}	
	Then the optimal meta-prameter $\Theta^*$ is obtained by maximizing the expectation over all query sets in the source environments as: 
	\begin{align}
	\Theta^* = \arg\max_{\Theta}\mathbb{E}_{\mathcal{D}^\mathrm{Q}} \left[ P(\Phi|\mathcal{D}^\mathrm{Q}) \right].
	\end{align}		
    Upon a new target environment with dataset $\hat{\mathcal{D}}$ (containing $\hat{\mathcal{D}}^\mathrm{S}$ as the support set with a minimum level of labeled data, and the rest unlabeled for final testing), $\Theta^*$ is transferred to this environment and the base network $f_\Phi$ is refined according to $\hat{\mathcal{D}}^\mathrm{S}$.
	%
	%
	The two major meta-learning approaches applicable in sensing context differ in how $P_\Theta(y|\mathbf{x}, \mathcal{D}^\mathrm{S})$ is modelled.
	%
	
	
	
	%
	%
	
	\subsubsection{Optimization-Based}
	This line of research focuses on the optimization-based training algorithms, aiming to cope with few observations or to converge within few optimization steps. Essentially, training algorithms are adjusted to find good initialization weights as $\Theta$, so that $f_\Phi$ could be generalized to new tasks. 
	These approaches model $P_\Theta(y|\mathbf{x}, \mathcal{D}^\mathrm{S})$ as $P_{g_{\Theta_g(\mathcal{D}^\mathrm{S})}}(y|x) $, where
	\begin{align}
	    g_{\Theta_g(\mathcal{D}^\mathrm{S})} &= 
	    g_{\Theta_g}\left(\Theta_0, \{\nabla_{\Theta_0}\mathcal{L}(\mathbf{x}_i,y_i)\}_{(\mathbf{x}_i,y_i)\in \mathcal{D}^\mathrm{S}}\right),
	\end{align}	
	is the \textit{meta learner} that generates initialization weights with the gradient of loss $\{\nabla_{\Theta_0}\mathcal{L}(\mathbf{x}_i,y_i)\}_{(\mathbf{x}_i,y_i)\in \mathcal{D}^\mathrm{S}}$ and starting weights $\Theta_0$ as input.
	Model-Agnostic Meta-Learning (MAML)~\cite{b26_maml} claims to be applicable to any network learnt via gradient descent, 
	%
    whereas later proposals (e.g., Reptile~\cite{nichol2018reptile}, Meta-SGD~\cite{meta_sgd}, TAML~\cite{meta_taml}) all intend to improve learning efficiency along various directions.
    %
    However, as stated in~\cite{sun2019meta,meta_leo}, the generalization ability of these approaches is questionable. To be specific, when they are applied to a base network $f_{\Phi}$ with a high-dimensional parameter space, tuning initialization weights $\Theta$ directly via $g_{\Theta_g}(D^\mathrm{S})$ could result in generalization difficulty given very few observations. 
    %
    Recently, MetaSense~\cite{gong2019metasense} adopts MAML to construct an adaptive wearable sensing system. Compared with RF sensing data, the dimension of wearable sensing data obtained by IMUs is much lower. Therefore, they adopt shallow neural network as $f_\Phi$ to avoid the generalization challenge described above. RF sensing (given the data model described in Section~\ref{ssec:model}) certainly demands a powerful base network $f_\Phi$ (with a high-dimensional parameter space for $\Theta$), rendering optimization-based meta-learning approaches largely inapplicable. 
    
    \subsubsection{Metric-Based} This category of non-parametric approaches intend to classify a new observation $\mathbf{x}$ by a weighted sum of the labels in $\mathcal{D}^\mathrm{S}$ as follows:
    \begin{align}
    \label{eq:metric}
        P_\Theta(y|\mathbf{x}, \mathcal{D}^\mathrm{S}) =  \sum\limits_{(\mathbf{x}_i,y_i)\in \mathcal{D}^\mathrm{S}} k_\Theta(f_\Phi(\mathbf{x}),f_\Phi(\mathbf{x}_i))y_i ,
    \end{align}
    where $k_\Theta$ is an optimal metric function for measuring the similarity between $\mathbf{x}$ and $\mathbf{x}_i$. Essentially, they aim to learn an embedding function $f_\Phi$ that transforms inputs into a representation suitable for classification via the similarity comparison. Although several proposals have been made under this category (e.g., Siamese networks~\cite{b23_siamese}, Matching networks~\cite{b18_matching_net}, Prototypical networks~\cite{b19_proto_net}), they mostly differ in the choice of the (input) embedding vectors and non-parametric distance metric (e.g., Cosine similarity).
    %
    %
    Compared with optimization-based approaches, the meta-training phase of metric-based approaches is rather straightforward: it mainly focuses on learning a powerful distance metric to achieve generalization rather than directly tuning $f_\Phi$. This property has made metric-based meta-learning approaches less constrained by the complexity of $f_\Phi$, and it has also motivated us to adopt metric-based approaches for achieving one-shot RF-HAR in this paper.

	\section{\systemname: One-Shot HAR}
	\label{sec:metaRF}
    Based on our discussions in Section~\ref{sec:bg}, we hereby present {\systemname} comprising two novel designs: i) a meta-learning framework that involves a parametric RF-specific module for training a powerful distance metric, and ii) a dual-path base network that fully exploits the high-dimensional features contained in the signal matrix (thus applicable to all three RF sensing datasets). We first describe the problem formulation. Then we elaborate our meta-learning framework and dual-path base HAR network. 
    

	\subsection{Problem Formulation}
	\label{sec:problem_formulate}
	In this paper, the ultimate goal of {\systemname} $q_\Omega$, parameterized by $\Omega$, is to perform one-shot RF-HAR, i.e., adapting to every new environment rapidly with a single labeled observation per class. To achieve it, we need a base HAR network $f_\Phi$, parameterized by $\Phi$, to \textcolor{black}{extract features} from input observations (i.e., RF signal matrix). As introduced in Section~\ref{ssec:model}, we deem an RF signal matrix $\mathbf{x}\in \mathbb{R}^{K \times L \times N_\mathrm{r}}$ as an image, where $K$ is slow time dimension, $L$ is fast time dimension, and $N_\mathrm{r}$ is the number of tx-rx pairs. This base network $f_\Phi$ is then wrapped into a meta-learning framework to be generalized to new environments. In order to maintain the generalization capability of $f_\Phi$ given its high-dimensional parameter space, we adopt a metric-based meta-learning framework $g_\Theta$ parameterized by $\Theta$. Therefore, {\systemname} $q_\Omega$ includes the base network $f_\Phi$ and meta-learning framework $g_\Theta$, hence $\Omega = \Phi \cup \Theta$.
	
	The procedure for learning $q_\Omega$ is planned as follows. We first train {\systemname} with $N_\mathrm{e}$ source environment training \textcolor{black}{datasets} $\mathcal{D} = \{\mathcal{D}_\ell\}_{\ell=1}^{N_\mathrm{e}}$. For each epoch, we train $q_\Omega$ on environment datasets, learning from environment to environment to mimic how {\systemname} would be tested when presenting in a new environment. 
	More specifically, for each environment dataset $\mathcal{D}_\ell$, we sample support observations as support set $\mathcal{D}_{\ell}^\mathrm{S} =  \{\mathbf{x}_{\ell,j}^\mathrm{S}, y_{\ell,j}^\mathrm{S}\}_{j=1}^{N_\mathrm{c}}$, $\mathbf{x}_{\ell,j}^\mathrm{S} \in \mathbb{R}^{N_\mathrm{s} \times K \times L \times N_\mathrm{r}}$ and query observations as query set $\mathcal{D}_{\ell}^\mathrm{Q} = \{\mathbf{x}_{\ell}^\mathrm{Q}, y_{\ell}^\mathrm{Q}\}$, $\mathbf{x}_{\ell}^\mathrm{Q} \in \mathbb{R}^{1 \times K \times L \times N_\mathrm{r}}$, where $N_\mathrm{s}$ denotes the number of observations (i.e., $N_\mathrm{s}$ = 1 for \textit{one-shot} learning), $N_\mathrm{c}$ is the number of activity categories, and $y_{\ell}$ denotes activity label. Note that the support set $\mathcal{D}^\mathrm{S}_\ell$ and query set $\mathcal{D}^\mathrm{Q}_\ell$ belong to the same environment space but observations are disjoint, i.e., $\mathcal{D}^\mathrm{S}_\ell \cap \mathcal{D}^\mathrm{Q}_\ell = \varnothing$. We drop the subscript $\ell$ as the learning procedure is identical in all environments. Essentially, the objective function of {\systemname} $q_\Omega$ can be formulated as follows:
	
	\begin{align}
	\Omega^* = \arg\max_{\Omega}\mathbb{E}_\mathcal{D} \left[\sum \limits_{\mathcal{D}^\mathrm{Q}} \sum \limits_{\mathcal{D}^\mathrm{S}} g_\Theta \left(f_\Phi\left(\mathbf{x}^{\mathrm{Q}}\right), f_\Phi\left(\mathbf{x}^{\mathrm{S}}\right),y^{\mathrm{S}}\right) \right].
	\end{align}

	\begin{figure*}[t]
	\centering
	\subfloat[{\systemname} $\boldsymbol{q_\Omega}$.]
	{
	    \begin{minipage}[b]{0.5\linewidth}
		\centering
		\includegraphics[width=.78\textwidth]{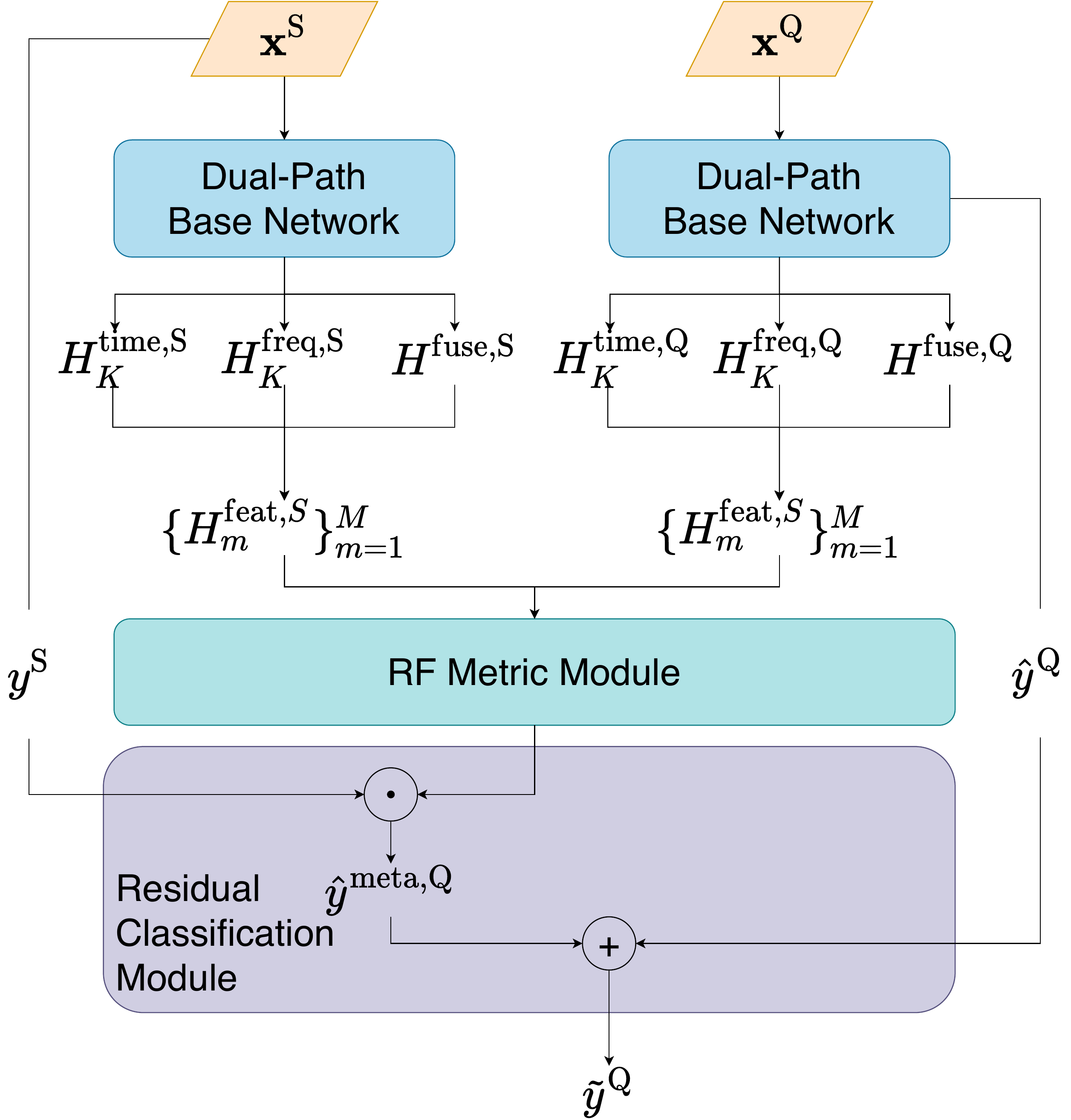}
		\end{minipage}
		\label{fig:meta_net}
	}
	\hfill
	\subfloat[Dual-path base network $\boldsymbol{f_\Phi}$.]
	{
	    \begin{minipage}[b]{0.48\linewidth}
		\centering
		\includegraphics[width=.78\textwidth]{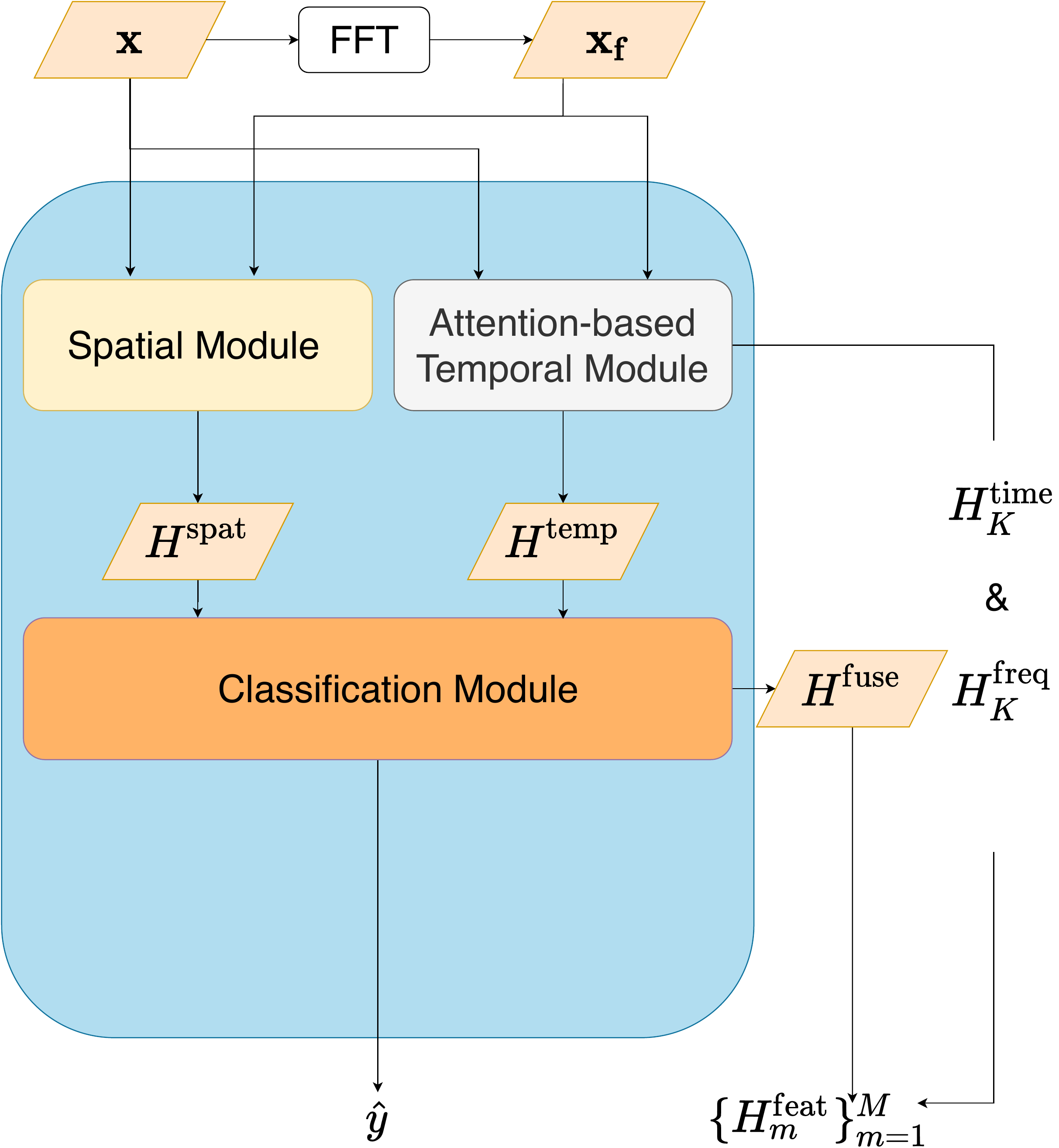}
		\end{minipage}
		\label{fig:dual}
	}
	\vspace{-1ex}
	\caption{Overall design of {\systemname} $\boldsymbol{q_\Omega}$ and its dual-path base network $\boldsymbol{f_\Phi}$.}
	\label{fig:overall}
    \end{figure*}
    
	\subsection{Meta Framework}
	\label{sec:meta-learning}
	We intend to design a trainable metric-based meta-learning framework specifically for RF signal matrices. As illustrated in Figure~\ref{fig:meta_net}, our proposed meta framework $g_\Theta$ consists of two modules:
	\begin{itemize}
	    \item RF Metric Module $g_{\mathrm{metric}}$: it aims to train a powerful distance metric for measuring observations similarities via a linear mapping layer. Meanwhile, it takes into consideration the multiple features
	    generated by the dual-path base network $f_\Phi$. Consequently, this module could provide a holistic interpretation of $\mathbf{x}$ toward better generalization.   
	    \item Residual Classification Module $g_{c}$: it intends to incorporate the capability of $f_\Phi$ further as a recognition network, in addition to its feature extractor role in $g_{\mathrm{metric}}$. This module allows the base network to assist in (meta)-training the distance metric $g_{\mathrm{metric}}$.
	    
	\end{itemize}
	
	
	
	\subsubsection{RF Metric Module} 
	\label{sssec:rf_metric}
	RF metric module first employs our dual-path base HAR network (see Section~\ref{sec:har}) as a (deep) feature extractor $f_{\Phi} : \mathbb{R}^{D_{ \mathbf{{x}}}}  \rightarrow \mathbb{R}^{D_{\mathbf{z}}} $, mapping an RF signal matrix $\mathbf{x}$ to a feature representation $\mathbf{z}$, where $D_{\mathbf{x}}$ and $D_{\mathbf{z}}$ are the dimensionalities of RF signal matrix and the corresponding embedding, respectively. Different from previous works such as~\cite{b18_matching_net,b14_csi,b15_csi}, we expect that the metric space of multiple features is able to capture a more complex representation than simply relying on single feature representation. Therefore, considering the RF signal matrix described in \textcolor{black}{both time and frequency domains}, we leverage $M$ features to represent the RF signal matrix $\mathbf{x}$ as follows:
	\begin{align}
	    \{ H_{m}^\mathrm{feat} = f_{\Phi} (\mathbf{x}) \}_{m=1}^{M}. \nonumber
	\end{align}
	In our problem setting, 
	$H_{m}^\mathrm{feat}$ includes $H^\mathrm{time}$ and $H^\mathrm{freq}$ respectively extracted from time and frequency domains, \textcolor{black}{as well as $H^\mathrm{fuse}$ combining} features from both domains to derive spatial and temporal information. We refer to Section~\ref{sec:har} for further elaborations on these features. Given these features, we employ cosine distance $d(a,b) = - \dfrac{a \cdot b}{\norm{a}\norm{b}}$ to obtain a distance set $\{\lambda_m\}_{m=1}^M$ between features generated by support observations and query observations as follows:
	\begin{align} 
	\{\lambda_m\}_{m=1}^M = \left\{d\left(H_{m}^\mathrm{feat,S}, H_{m}^\mathrm{feat,Q}\right)\right\}_{m=1}^{M}. \nonumber
	\end{align}
	In order to further increase the RF metric space, we design RF metric module to be trainable instead of solely relying on cosine distance. We stack these $M$ distances $[\lambda_1,\lambda_2,...,\lambda_M]$ into a vector form $\mathbf{\Lambda}$. \textcolor{black}{Given a query observation}, we propose to combine each distance \textcolor{black}{measure} $\lambda_m$ in $\mathbf{\Lambda}$ via learnable linear mapping weights $\bm{\eta}$. Therefore, we could compute the activity probabilities of query observations by weighting the labels of support observations $y^\mathrm{S}$ as follows:
	\begin{align}\label{eq:softmax}
	p_{\bm{\beta}}(y=j|\bm{x}) = \text{softmax} (- \Lambda_j \bm{\beta}), \bm{\beta} = \bm{\eta}  y^\mathrm{S},
	\end{align}
	where $\bm{\eta} \in \mathbb{R}^{M \times N_\mathrm{c}}$ and $\Lambda_j \in \mathbb{R}^{1 \times M}$.
	To further realize the role of $\bm{\beta}$, we analyze that for $j$-th class, the class-wise cross-entropy loss function is given by
	\begin{align} \label{eq:loss}
	L_j(\bm{\beta}) = \sum_{\bm{x} \in \mathcal{D}^\mathrm{Q} } \left [ \Lambda_j \bm{\beta} + \log  \left ( \sum_{n=1}^{N_\mathrm{c}} e^{ -\Lambda_n \bm{\beta}}  \right )  \right ].
	\end{align}
	For each sample of query set $\bm{x} \in \mathcal{D}^\mathrm{Q}$, the second-order partial derivative of Eq~\eqref{eq:loss} with respect to $\bm{\beta}$ is
	\begin{align}
	    \nabla^2 L_j(\bm{\beta}) = \frac{1}{ \bm{1}^T \gamma } \text{diag} (\gamma) - \frac{1}{(\bm{1}^T \gamma)^2} \gamma \gamma^T,
	\end{align}
	where $\gamma = [ \gamma_1, \cdots, \gamma_{N_c} ]$ and  $\gamma_k = e^{- \Lambda_k \bm{\beta}}$. If $ \triangledown^2 L_j(\bm{\beta}) \geq 0 $, $L_j(\bm{\beta})$ in Eq~\eqref{eq:loss} is convex. Therefore, we need to verify that $v^T \triangledown^2 L_j(\bm{\beta})  v \geq 0$ for all $v$, but we have:
	\begin{align}
	&v^T \nabla  ^2 L_j(\bm{\beta})  v &=
	 \frac{(\sum_{n=1}^{N_\mathrm{c}} \gamma_n v_n^2 ) (\sum_{n=1}^{N_\mathrm{c}} \gamma_n  ) - ( \sum_{n=1}^{N_\mathrm{c}} v_n \gamma_n )^2 }{ ( \sum_{n=1}^{N_\mathrm{c}} \gamma_n  )^2 }, \nonumber
	\end{align}
	which is indeed non-negative due to Cauchy-Schwarz inequality, i.e.,  $(\sum_{n=1}^{N_\mathrm{c}} v_n\gamma_n  )^2 \leq ( \sum_{n=1}^{N_\mathrm{c}} \gamma_n v_n^2 ) ( \sum_{n=1}^{N_\mathrm{c}} \gamma_n  ) $. Now we have proven the convexity of $L_j(\bm{\beta})$, which in turn indicates that our RF metric module can quickly learn the optimal parameters $\bm{\beta}$ for combining multiple representations together.

	
	
	
	\subsubsection{Residual Classification Module} 
	In this module, different from traditional metric-based meta-learning framework, we aim to employ the base network $f_\Phi$ for recognition too, rather than solely as a feature extractor explained in Section~\ref{sssec:rf_metric}. To be specific, we first compute $\hat{y}^{\mathrm{meta,Q}}$, i.e., classify query observations by weighting the labels of support observation $y^\mathrm{S}$, as described in Eq~\eqref{eq:softmax}. Meanwhile, we exploit $f_\Phi$ as recognition network to compute logits directly $\hat{y}^\mathrm{Q} = f_\Phi(\mathbf{{x}}^\mathrm{Q})$. Afterwards, we incorporate $\hat{y}^\mathrm{Q}$ into $\hat{y}^{\mathrm{meta,Q}}$ via a residual connection, so that the final predicted logits of {\systemname} \textcolor{black}{$\tilde{y}^\mathrm{Q}$} is computed as $ g_c(g_{\mathrm{metric}}(f_\Phi(\mathbf{{x}}^\mathrm{S}),f_\Phi(\mathbf{{x}}^\mathrm{Q})),y^\mathrm{S},\hat{y}^\mathrm{Q}) = g_{\mathrm{metric}}(f_\Phi(\mathbf{{x}}^\mathrm{S}),f_\Phi(\mathbf{{x}}^\mathrm{Q}))y^\mathrm{S} + \hat{y}^\mathrm{Q}$. \textcolor{black}{It enables the base network to reinforce the meta-training of} the distance metric $g_{\mathrm{metric}}$.
	%
    %
    %
	\subsubsection{Training Strategy}
	We carefully devise our training strategy for learning {\systemname}. As described in \textbf{Algorithm~\ref{algo}}, given {\systemname} $q_\Omega$ including the base network $f_\Phi$ and meta-learning framework $g_\Theta$, we exploit training dataset $\mathcal{D}$ and testing dataset $\hat{\mathcal{D}}$ in training and testing stages, respectively. 
		\begin{algorithm}[b]
		\caption{\systemname\ training.}
		\begin{algorithmic}[1]
			\REQUIRE Training dataset $\mathcal{D}$, Testing dataset $\hat{\mathcal{D}}$, Base network $f_{\Phi}$, {\systemname} $q_{\Omega}$, Meta-learning framework $g_{\Theta}$, hyperparameters $\alpha$ and $\beta$, as well as the maximum number of iterations $\mathrm{iter}$
			\STATE \textsf{\% Training}
			\WHILE{$\mathrm{iter}~!= 0$}
			\label{line:train_start}
			\STATE Sample environment minibatch $\{\mathcal{D}_\ell\} \sim \mathcal{D}$ 
			\FOR {$\mathcal{D}_\ell \in \{\mathcal{D}_\ell\}$}
			\STATE Sample support and query observations $\mathcal{D}_{\ell}^\mathrm{S}, \mathcal{D}_{\ell}^\mathrm{Q} \sim \mathcal{D}_{\ell}$ 
			\STATE $\mathcal{D}_{\ell}^\mathrm{S} =  \{\mathbf{x}_{\ell,j}^\mathrm{S}, y_{\ell,j}^\mathrm{S}\}_{j=1}^{N_\mathrm{c}}$, $\mathcal{D}_{\ell}^\mathrm{Q} =   \{\mathbf{x}_{\ell}^\mathrm{Q}, y_{\ell}^\mathrm{Q}\}$, $\mathcal{D}_{\ell}^\mathrm{Q} \cap \mathcal{D}_{\ell}^\mathrm{S} = \varnothing $
			\STATE Evaluate $\nabla_{\Omega} \mathcal{L}_\ell(f_{\Phi}(\mathbf{x}_{\ell}^\mathrm{S}), y^\mathrm{S}_\ell)$,
			\label{line:ft_start} 
			~~$\Omega \leftarrow \Omega - \alpha\nabla_{\Omega} \mathcal{L}_\ell$
			\STATE Evaluate $\nabla_{\Theta} \mathcal{L}_\ell(q_{\Omega}(\mathbf{x}_{\ell}^\mathrm{S}, \mathbf{x}_{\ell}^\mathrm{Q}, y^\mathrm{S}_\ell), y^\mathrm{Q}_\ell)$, 
			\label{line:meta_train_start}	
			~~$\Theta \leftarrow \Theta- \beta \nabla_{\Theta}\mathcal{L}_\ell$
			%
			\ENDFOR 
			\STATE $\mathrm{iter} \leftarrow \mathrm{iter} -1$
			\ENDWHILE
			\label{line:train_end}			
			\STATE \textsf{\% Testing}
			\FOR{$\hat{\mathcal{D}_\ell} \in \hat{\mathcal{D}} $}
			\STATE Sample support and query observations $\hat{\mathcal{D}}_{\ell}^\mathrm{S}, \hat{\mathcal{D}}_{\ell}^\mathrm{Q} \sim \hat{\mathcal{D}}_{\ell}$ 
			\STATE $\hat{\mathcal{D}}_{\ell}^\mathrm{S} =  \{\mathbf{x}_{\ell,j}^\mathrm{S}, y_{\ell,j}^\mathrm{S}\}_{j=1}^{N_c}$, $\hat{\mathcal{D}_{\ell}}^\mathrm{Q} =   \{\mathbf{x}_{\ell}^\mathrm{Q}, y_{\ell}^\mathrm{Q}\}$, $\hat{\mathcal{D}}_{\ell}^\mathrm{Q} \cap \hat{\mathcal{D}}_{\ell}^\mathrm{S} = \varnothing $ 
			\STATE Evaluate $\nabla_{\Theta} \mathcal{L}_\ell(f_{\Phi}(\mathbf{x}_{\ell}^\mathrm{S}), y^\mathrm{S}_\ell)$,
			~~$\Theta \leftarrow \Theta - \alpha\nabla_{\Theta}\mathcal{L}_\ell$ \label{line:ftune}
			%
			\STATE Predict 	$\textcolor{black}{\tilde{y}^\mathrm{Q}} = q_{\Omega}(\mathbf{x}_{\ell}^\mathrm{S}, \mathbf{x}_{\ell}^\mathrm{Q},\mathbf{y}_{\ell}^\mathrm{S})$
			\ENDFOR
		\end{algorithmic}
		\label{algo}
	\end{algorithm}		
    In training stage (line \ref{line:train_start}-\ref{line:train_end}), 
    we train $f_\Phi$ and $g_\Theta$ in the leader-follower asymmetric manner between inner- and meta-training to enhance the overall capability on performing one-shot RF-HAR with $q_\Omega$. More specifically, we first inner train $f_\Phi$ with $\mathcal{D}^\mathrm{S}$ (line~\ref{line:ft_start}). 
    We minimize the inner training loss $\mathcal{L}(f_{\Phi}(\mathbf{x}^\mathrm{S}),
    y^\mathrm{S})$ by conducting gradient descent with respect to both base network parameters $\Phi$ and meta parameters $\Theta$. This helps finding a coarse initialization points for learning meta parameters $\Theta$, boosting the convergence speed of \systemname\ $q_\Omega$. Afterwards (line \ref{line:meta_train_start}), we meta-train $g_\Theta$ conditioned on learned $f_\Phi$ using both support and query observations, so that it produces $q_\Omega$ that performs well on recognizing query observations. 

    Now we have achieved a well-trained RF-HAR $q_\Omega$ by sequentially training $f_\Phi$ and $g_\Theta$. Given a testing dataset $\hat{\mathcal{D}}$, we could evaluate the performance of {\systemname} on performing one-shot RF-HAR. We first intend to refine $q_\Omega$ with labeled support observations $\hat{\mathcal{D}}^\mathrm{S} =  \{\mathbf{x}_{j}^\mathrm{S}, y_{j}^\mathrm{S}\}_{j=1}^{N_c}$. To be specific, we adapt the well-trained $q_\Omega$ using the same inner training procedure, except that we only fine-tune meta-learning framework parameter $\Theta$ (lines~\ref{line:ftune}). 
    Finally, we classify query observations with refined $q_\Omega$ and labeled support observations as \textcolor{black}{$\tilde{y}^\mathrm{Q} = q_{\Omega}(\mathbf{x}^\mathrm{S}, \mathbf{x}^\mathrm{Q},\mathbf{y}^\mathrm{S})$.}

	\subsection{Dual-Path Base Network}\label{sec:har}
	As illustrated in Figure~\ref{fig:dual}, our base HAR network \textcolor{black}{$f_{\Phi}$} is composed of three main modules to learn RF feature representations:
	\begin{itemize}
		\item Spatial Module $f_\mathrm{s}$: it \textcolor{black}{extracts}
		sensitive spatial features from \textcolor{black}{RF matrices $\mathbf{x}$} in both time and frequency domains.
		\item Attention-based Temporal Module $f_\mathrm{t}$: it aims to capture long-term temporal features from RF matrices in both time and frequency domains. And it also intends to generate joint \textcolor{black}{temporal representations across two domains.} 
		\item Classification Module $f_\mathrm{c}$: this final module predicts activity label, given trained features from $f_\mathrm{s}$ and $f_\mathrm{t}$.
	\end{itemize}
	%
    %
	For both $f_\mathrm{s}$ and $f_\mathrm{t}$, we first compute $\mathbf{x_f} \in \mathbb{R}^{K\times L\times N_\mathrm{r}}$ via FFT along the slow time of an RF matrix $\mathbf{{x}} \in \mathbb{R}^{K\times L\times N_\mathrm{r}}$. Then both $\mathbf{x}$ (time domain) and $\mathbf{x_f}$ (frequency domain) are used as input.

	\begin{figure}[b]%
        \vspace{-1ex}
		\centering
		\subfloat[Spatial Separate Module]{\includegraphics[width = 0.21\textwidth]{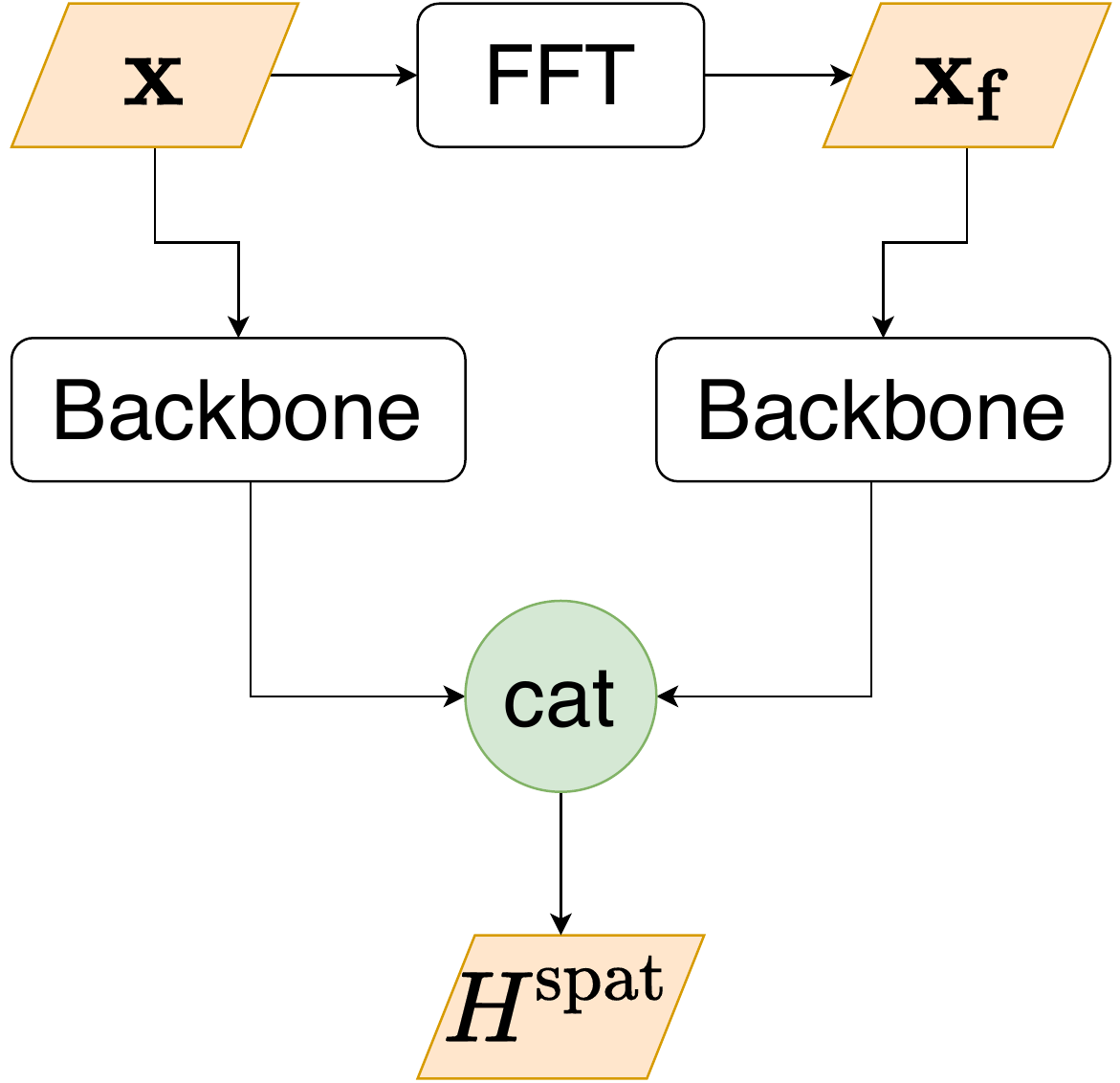}
			\label{fig:sp_sep}}\qquad
		\subfloat[Spatial Fuse Module]{\includegraphics[width = 0.21\textwidth]{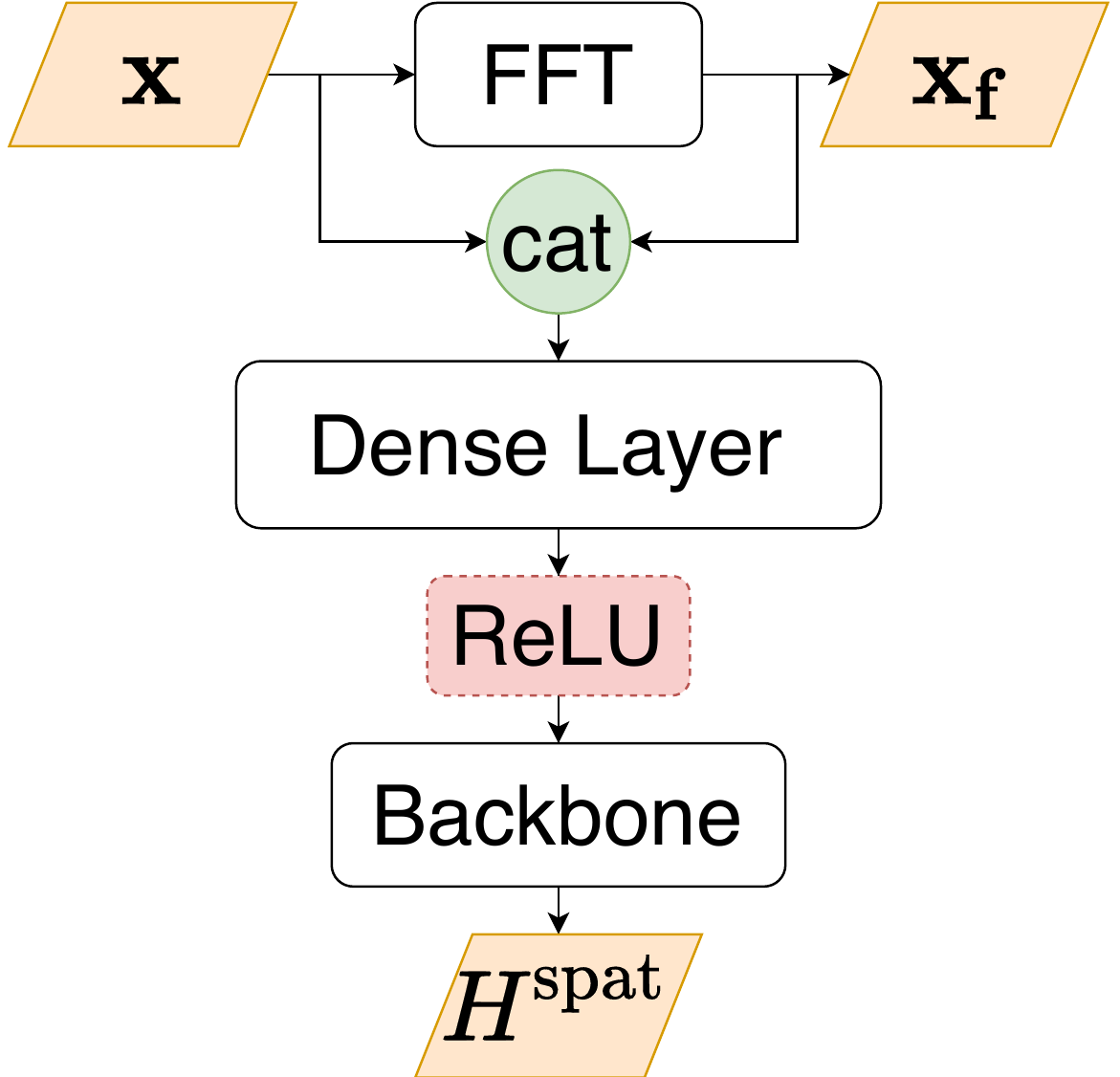}
			\label{fig:sp_fuse}
		}
	    \vspace{-1ex}		
		\caption{Two alternative spatial modules.}
		\label{spatial}
	\end{figure}		
	
	\subsubsection{Spatial Module} This module leverages $f_\mathrm{s}$ to learn the spatial features $H^\mathrm{spat}$. Essentially, \textcolor{black}{it extracts} sensitive spatial features from RF matrices in both time domain $\mathbf{x}$ and frequency domain $\mathbf{x_f}$. We employ state-of-the-art deep learning model CNNs as the backbone $f_\mathrm{b}$, \textcolor{black}{and we empirically select the most cost-effective backbone in Section~\ref{sssec:backbone}.} Generally, CNNs are designed for exploring spatial features in image data. Therefore, we regard $\mathbf{x}$ and $\mathbf{x_f}$ as image with $N_\mathrm{r}$ tx-rx pair as input channels, $L$ fast time dimension as height of input planes in pixels and $K$ slow time dimension as width in pixels. Moreover, we additionally employ one convolutional layer $f_\mathrm{a}$ to adjust channel dimension into designated dimension declared by the backbone.

	We explore two spatial modules, named spatial separate module and spatial fuse module. As depicted in Figure~\ref{fig:sp_sep}, spatial separate module aims to employ backbone for generating spatial features of $\mathbf{x}$ and $\mathbf{x_f}$ \textcolor{black}{separately:} $H^{\mathrm{s}, \mathbf{x}} = f_\mathrm{b}(f_\mathrm{a}(\mathbf{x}))$ and $H^{\mathrm{s}, \mathbf{x_f}} = f_\mathrm{b}(f_\mathrm{a}(\mathbf{x_f}))$; \textcolor{black}{the final spatial features $H^\mathrm{spat}$ are then produced} by concatenating $H^{\mathrm{s}, \mathbf{x}}$ and $H^{\mathrm{s}, \mathbf{x_f}}$. However, spatial separate module involves a high computation complexity due to repeated backbone usage. Therefore, we further explore spatial fuse module illustrated in Figure~\ref{fig:sp_fuse}. \textcolor{black}{As a simplification, we concatenate} $\mathbf{x}$ and $\mathbf{x_f}$ at the initial stage $\mathbf{x_c} = [\mathbf{x},\mathbf{x_f}]\in \mathbb{R}^{K\times L \times 2N_\mathrm{r}}$. Afterwards, we intend to learn a composite representation from $\mathbf{x_c}$. We first reshape $\mathbf{x_c}\in \mathbb{R}^{K\times 2N_\mathrm{r}L}$ and then we employ a \textit{Rectified Linear Unit} (ReLU) activated dense layer \textcolor{black}{to generate} a composite representation $H^\mathrm{c} \in \mathbb{R}^{K\times\alpha\times2 }$, where $\alpha$ denotes the hidden dimension. Afterwards, we employ backbone $f_\mathrm{b}$ to directly extract $H^\mathrm{spat} = f_\mathrm{b}(f_\mathrm{a}(H^\mathrm{c}))$. The corresponding performance comparisons between these two modules, \textcolor{black}{along with the choice of $\alpha$ and activation function}, are conducted in Section~\ref{sec:eval}.


	\begin{figure}[t]
		\centering
		\includegraphics[width = 0.46\textwidth]{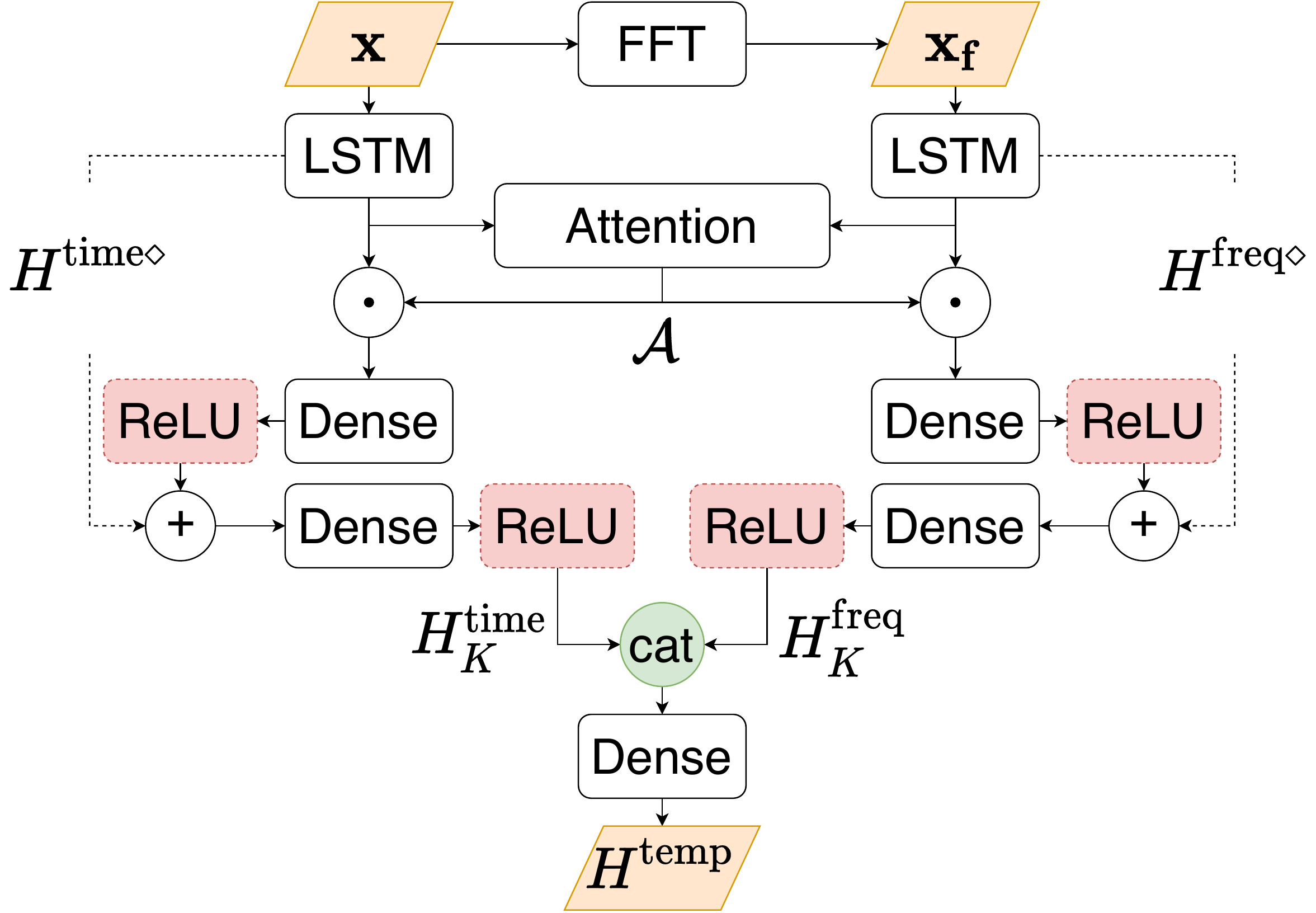}
	    \vspace{-1ex}		
		\caption{Attention-based temporal module.}
		\label{fig:temporal module}
		\vspace{-2ex}
	\end{figure}	
	
	\subsubsection{Attention-Based Temporal Module}
	In order to generate long-term temporal features from RF matrices in both time and frequency domains, we utilize $f_\mathrm{t}$ to learn \textcolor{black}{attended time features $H^\mathrm{time}$ and attended frequency features $H^\mathrm{freq}$}. As illustrated in Figure~\ref{fig:temporal module}, we first extract initial time features $H^\mathrm{time\diamond}$ and frequency features $H^\mathrm{freq\diamond}$, \textcolor{black}{leveraging} Long Short Term Memory (LSTM) to avoid gradient vanishing problems. Therefore, we reshape $\mathbf{x}\in\mathbb{R}^{K\times(L \times N_\mathrm{r})}$ and $\mathbf{x_f}\in\mathbb{R}^{K\times(L \times N_\mathrm{r})}$, \textcolor{black}{regarding the slow time dimension as step index with cardinality $K$}. Then, we extract $H^\mathrm{time\diamond} \in \mathbb{R}^{K \times \alpha}$ and $H^\mathrm{freq\diamond}\in \mathbb{R}^{K \times \alpha}$ by passing $\mathbf{x}$ and $\mathbf{x_f}$ into LSTM separately. After that, we focus on employing attention mechanism~\cite{kim2018bilinear} to learn richer joint representations between initial feature representations, \textcolor{black}{aiming to generate the attention map $\mathcal{A} \in \mathbb{R}^{K \times K}$ containing the joint information of every step between $H^\mathrm{time\diamond}$ and $H^\mathrm{freq\diamond}$. To be specific,
	ReLU activated dense layers are first employed} to map $H^\mathrm{time\diamond}$, $H^\mathrm{freq\diamond}$ into ${H^\mathrm{time\dagger}}$, ${H^\mathrm{freq\dagger}\in \mathbb{R}^{K \times \iota}}$, where $\iota$ denotes hidden dimension. Then the attention map is generated as $\mathcal{A}:= \mathrm{softmax}(W \circ {H^\mathrm{time\dagger}}({H^\mathrm{freq\dagger}})^{T})$,	where $\circ$ stands for Hadamard product, $T$ denotes transpose, and $W\in \mathbb{R}^{K \times \iota}$ denotes \textcolor{black}{weight} matrix. 
	
	\textcolor{black}{Combing the attention map $\mathcal{A}$ with the initial features produces}
	joint representations $((H^\mathrm{time\diamond})^T\mathcal{A})^T$ and $((H^\mathrm{freq\diamond})^T\mathcal{A})^T$, \textcolor{black}{and these joint representations are passed} through ReLU activated dense layers for adding non-linearity. Especially, we separately incorporate these mapped joint representations into $H^\mathrm{time\diamond}$, $H^\mathrm{freq\diamond}$ via a residual connection to generate $H^\mathrm{time}\in \mathbb{R}^{K \times \alpha}$, $H^\mathrm{freq} \in \mathbb{R}^{K \times \alpha}$, \textcolor{black}{thus it is able to solve vanishing gradients issue and achieve optimal identity mapping~\cite{he2016deep}.} Subsequently, we choose final step ($K$-th) of attended time and frequency features, i.e., $H^\mathrm{time}_K$ and $H^\mathrm{freq}_K$, for producing a composed temporal features $H^\mathrm{temp}$. To be specific, we first employ two ReLU activated dense layers to \textcolor{black}{generate} $H^\mathrm{time}_K$ and  $H^\mathrm{freq}_K$. Then, we stack those mapped features followed by a dense layer to compute $H^\mathrm{temp}\in \mathbb{R} ^ {1\times 2\alpha}$.
	
	\subsubsection{Classification Module} We have employed spatial module $f_\mathrm{s}$ and attention-based temporal module $f_\mathrm{t}$ to generate sensitive spatial features $H^\mathrm{spat}$ and attended temporal features $H^\mathrm{temp}$ from both time and frequency domains. In order to efficiently predict an activity label, we further derive an integrated feature by fusing both spatial and temporal representations. In this paper, we directly add spatial features $H^\mathrm{spat}$ and temporal features $H^\mathrm{temp}$ as $H^{\mathrm{fuse}} = H^\mathrm{temp} + H^\mathrm{spat}$. \textcolor{black}{To enable this integration, we set the hidden dimension $\alpha$ properly so as to equalize}
	the dimensions of $H^\mathrm{spat}$ and $H^\mathrm{temp}$. Finally, we utilize a dense layer to predict activity label \textcolor{black}{$\hat{y} = H^{\mathrm{fuse}}W_1$, where $W_1 \in \mathbb{R}^{2\alpha \times N_c}$ is weight matrix.}

	\vspace{1ex}
	\noindent\textbf{Remark}: We choose \textcolor{black}{the final step ($K$-th) of the attended temporal features $H^\mathrm{time}_K$ and $H^\mathrm{freq}_K$, along with the fused features $H^{\mathrm{fuse}}$, as $\{H_{m}^\mathrm{feat}\}_{m=1}^{M}$ in {\systemname} (see Section~\ref{sssec:rf_metric}).}

	\section{Experiments}\label{sec:pre}
	\label{sec:eval}
	In this section, we conduct extensive experiments on all three RF sensing techniques, i.e., Wi-Fi, FMCW, and impulse radio (IR), aiming to demonstrate the efficacy of the dual-path base network for activity recognition and to evaluate the overall performance of {\systemname} on performing one-shot RF-HAR. In particular, we report experiments on the following aspects: i) hyperparameter searching of network setting, ii) superiority of {\systemname} and dual-path base network over baselines, and iii) efficacy of {\systemname} and dual-path base network on various RF sensing datasets.
	
	\subsection{Datasets}
	As described in Section~\ref{ssec:model}, RF signal in each sensing technique could be formulated as matrix $\mathbf{x} \in \mathbb{R}^{K \times \ L \times N_r}$. In this section, we first elaborate RF signal matrix of these RF sensing techniques. Then we describe environment information and human activities provided in each dataset. 
	Given the variety of testing conditions, the activities of testing subjects may vary across different datasets.
	Details of each dataset is summarized in Table~\ref{tab:data_info}: for each RF sensing technology, we specify the number of environments involved, \textcolor{black}{the number of observations per environment taken,} and the number of activities tested. To generate different environments, \textcolor{black}{we first select a few rooms with distinct sizes, also involve different testing subjects. Within the same room, we change positions of the furniture and appliances, as well as the location of the subject. In order to artificially create ``differences'', we make sure that at least five objects (including the subject) have their position changed in a room when generating a new environment.} 
	Due to the popularity of Wi-Fi sensing, we choose to employ three datasets for it, but only one dataset for FMCW and IR. 
	\begin{table}[t]
	\caption{Datasets information.}
    \vspace{-1ex}
\begin{tabular}{|l|c|c|c|}
\hline
Sensing & Environments \# &  Observations \# & Activities \# \\\hline
Wi-Fi          & 80         & 25                           & 6          \\ \hline
Wi-Fi         & 100        & 20                           & 6          \\ \hline
Wi-Fi         & 120        & 16                           & 6          \\ \hline
FMCW        & 10         & 17                           & 6          \\ \hline
IR       & 50         & 16                           & 6          \\ \hline 
\end{tabular}
\label{tab:data_info}
\end{table}

	\subsubsection{Wi-Fi} 
	We exploit CSI of 30 OFDM subcarriers with 2 tx-rx pairs to record six human activities: wiping, walking, moving, rotating, sitting, and standing up. CSI information is sampled at 100~\!Hz and conducted window slicing size in 5.12~\!s. Therefore, each signal matrix is with a total $K = 512$ received packets, $L = 30$ subcarriers, and $N_r = 2$ tx-rx pairs. We employ 11 subjects and record from 6 different rooms. In the following experiments, we intend to investigate the impact of the number of environments as well. \textcolor{black}{Therefore, we split these environments data into three datasets including 80 environments, 100 environments, and 120 environments.} As described in Table~\ref{tab:data_info}, these datasets have 25, 20, and 16 observations per environment per activity, respectively.

	\subsubsection{FMCW} 
	We utilize FMCW radar device with 1 tx-rx pair to collect data. To be specific, we collect each frame with 253 frequency components every 67\!~ms. Then, we stack 100 frames for covering a 6.7~\!s interval. Therefore, each FMCW matrix $\mathbf{x} \in \mathbb{R}^{K \times \ L \times N_r}$ is with a total $K = 100$ frames, $L = 253$ frequency components, and $N_r = 1$ tx-rx pair. These matrices depict that 9 subjects in 2 different rooms perform six activities: standing up, sitting down, going out, entering room, putting on clothes, and putting off clothes. We aim to keep consistency on the number of observations per environment per activity of all datasets. Therefore, we split observations into 10 different environments.

	\subsubsection{IR} 
We employ IR device with 1 tx-rx pair to transmit pulse signal. Specifically, we collect each frame with 138 time components every 2.5\!~ms and we stack 400 frames together. Accordingly, each IR matrix $\mathbf{x} \in \mathbb{R}^{K \times \ L \times N_r}$ is with total $K = 400$ frames, $L = 138$ samples of a pulse, and $N_r = 1$ tx-rx pair. We employ 20 subjects in 3 rooms to perform six activities : sitting down, standing up, walking, falling, bending, and lying. We split all the observations into 50 different environments.

	\subsection{Baselines \& Implementation}
	We present several baseline schemes against which our \systemname\ and its base network will be compared. We also briefly explain the implementation of our experiments.
	
	\subsubsection{Baselines For Dual-path Base Network}
	We compare our proposed base network with the following baseline networks. We aim to demonstrate the efficacy of our base network on extracting features from RF signal matrix for HAR. 
	\begin{itemize}
		\item TIME: the proposed spatial module with only time domain data. It is used to evaluate network with spatial features in time domain.
		\item FREQ: the proposed spatial module with only frequency domain data. It is used to investigate the performance of network with spatial features in frequency domain.
		\item SPATIAL-SEP: the proposed spatial separate module. It is used to investigate the performance of network with spatial features in both time and frequency domain.
		\item SPATIAL-FUSE: the proposed spatial fuse module. It is used to examine network with spatial features in both time and frequency domain.
		\item DUAL-wo-ATT: the proposed dual-path base network without attention mechanism. It is used to demonstrate the efficacy of the attention mechanism. 
		\item DUAL-PATH: the proposed dual-path base network.
	\end{itemize}
	Note that, as SPATIAL-SEP and SPATIAL-FUSE are two alternative components, DUAL-wo-ATT and	DUAL-PATH will choose to involve whichever performs better.
	
	\subsubsection{Baselines For {\systemname}} 
	We evaluate the performance of one-shot human activity recognition when {\systemname} is adapted to new environments. Therefore, we mainly compare our network with the following meta-learning baselines and fine-tuning baseline.
	\begin{itemize}
		\item Fine-Tuning (FT): FT firstly trains dual-path base network with the training set. When FT adapts the network on the test set, the parameters of classification module would be fine-tuned with support observations on the test set. 
		\item MAML: a state-of-the-art optimization-based meta-learning baseline. \textcolor{black}{This scheme has been adopted by Metasense~\cite{gong2019metasense} to enable adaptive wearable sensing system}. MAML relies on meta-optimization through gradient descent in a model agnostic way. It expects to learn an initial representation that can be fine-tuned efficiently in a few steps. 
		\item Prototypical Network (PN): a state-of-the-art metric-based meta-learning baseline. Given few observations, PN generates prototypes (feature representations) for each class and it uses the Euclidean distance metric to search the closest prototype as predicted class.  
		\item {\systemname}*: our proposed meta-learning framework without the RF metric module.
		\item {\systemname}: our proposal. The comparison between {\systemname} and {\systemname}*  highlights the impact of RF metric module. 
	\end{itemize}
	\subsubsection{Implementation}
	We first apply data normalization to all datasets. Afterwards, we sample 80\% environments of each dataset into training dataset $\mathcal{D}$ and the remaining into testing dataset $\hat{\mathcal{D}}$, \textcolor{black}{and we perform a 10-fold cross-validation for each experiment.} We report the average accuracy evaluated on the testing dataset under 1-, 2-, 3-shot scenarios, i.e., $N_s = \{1,2,3\}$. We use regularization technique batch normalization~\cite{43442} to avoid overfitting. The following experiments are all programmed with PyTorch~\cite{NEURIPS2019_9015} and run on NVIDIA TESLA V100 with 16GB memory.    

	\begin{figure}[b]
		\centering
	    \includegraphics[width = 0.43\textwidth]{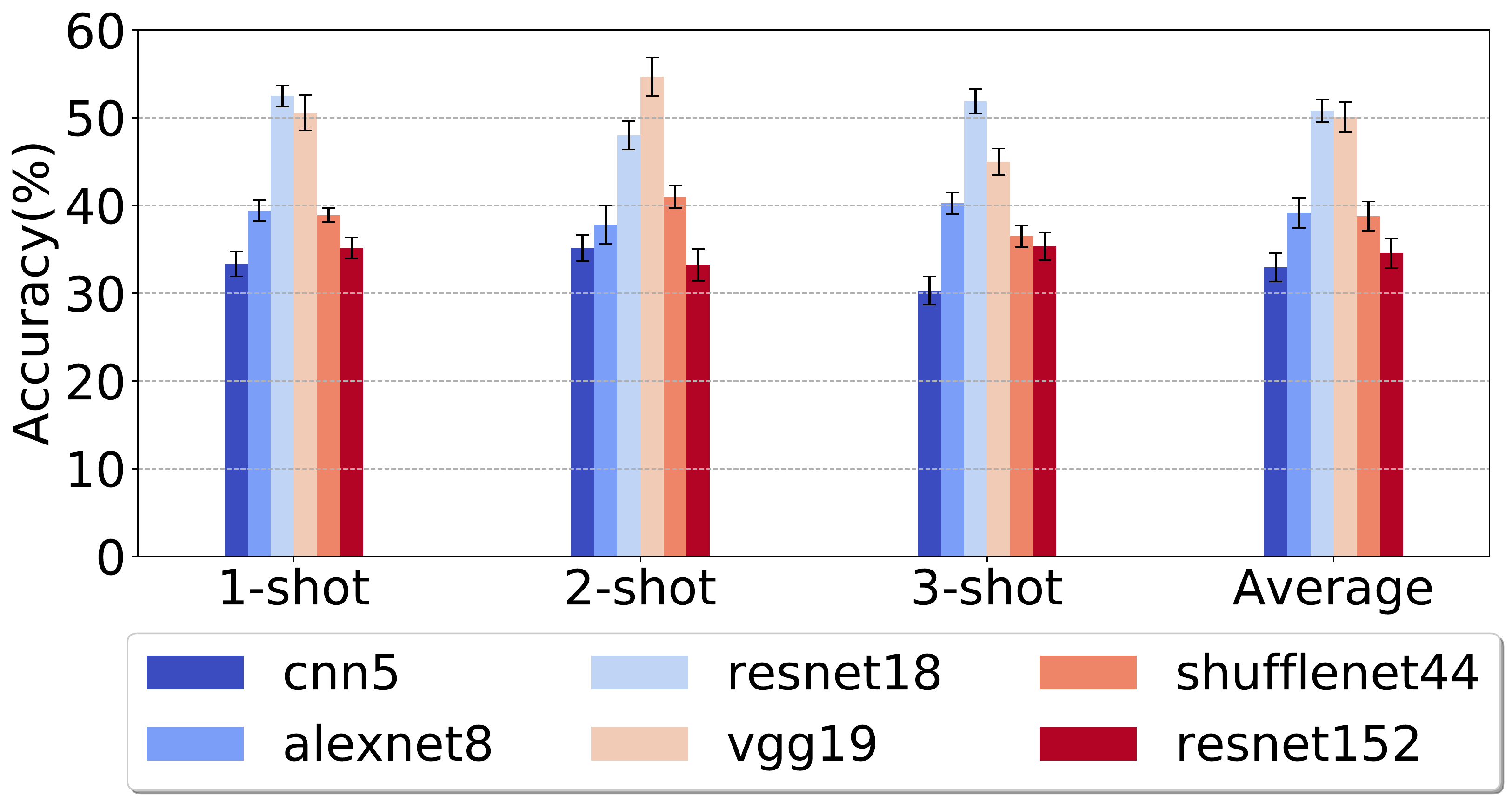}
	    \vspace{-1ex}		
		\caption{CNN backbone selection.}
		\label{fig:backbone}
	\end{figure}
	
	\begin{figure}[b]
		\centering
	    \includegraphics[width = 0.43\textwidth]{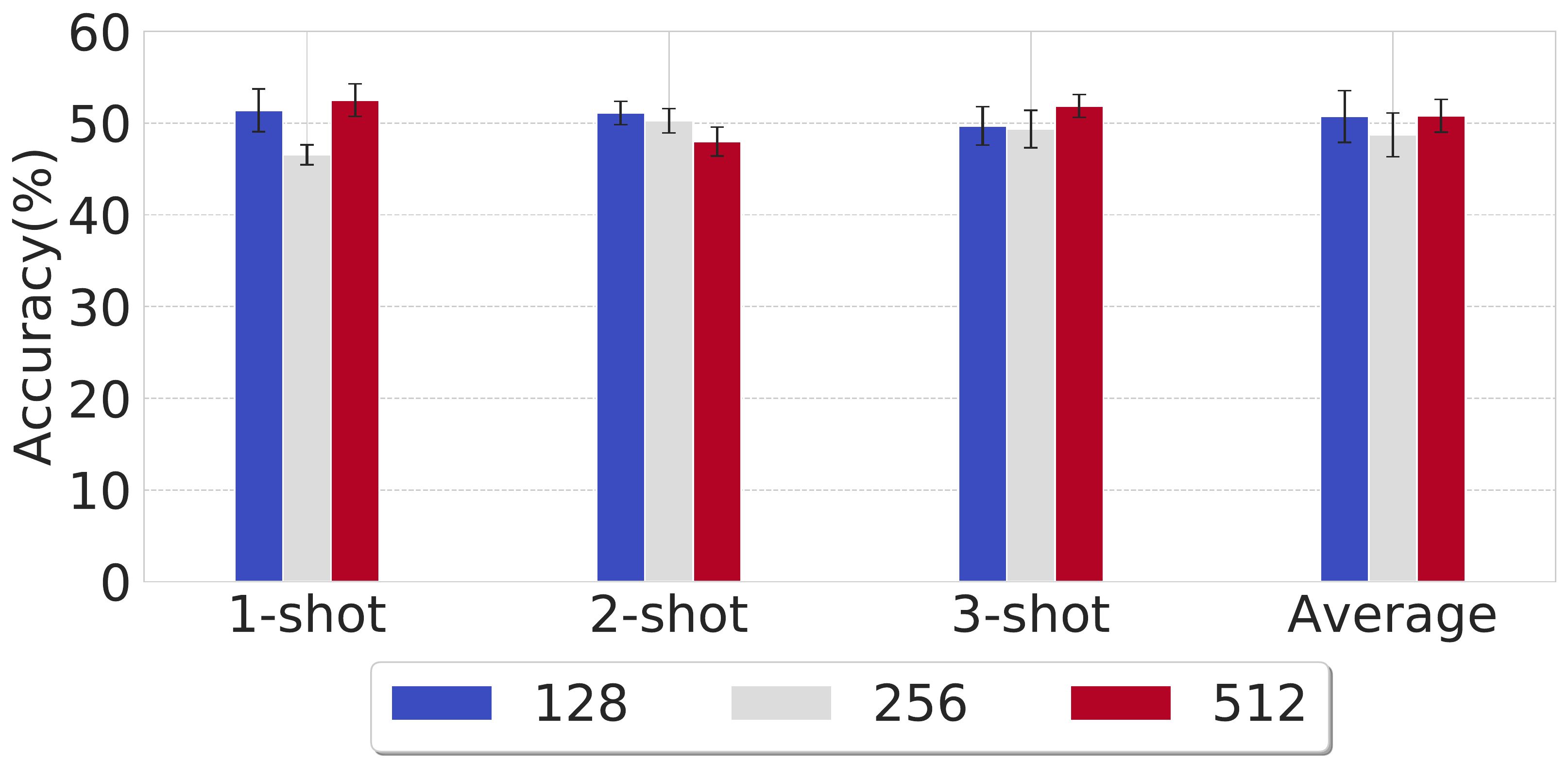}
	    \vspace{-1ex}		
		\caption{Hidden dimension size.}
		\label{fig:hidden_dim}
	\end{figure}
	
	\begin{figure}[b]
		\centering
	    \includegraphics[width = 0.43\textwidth]{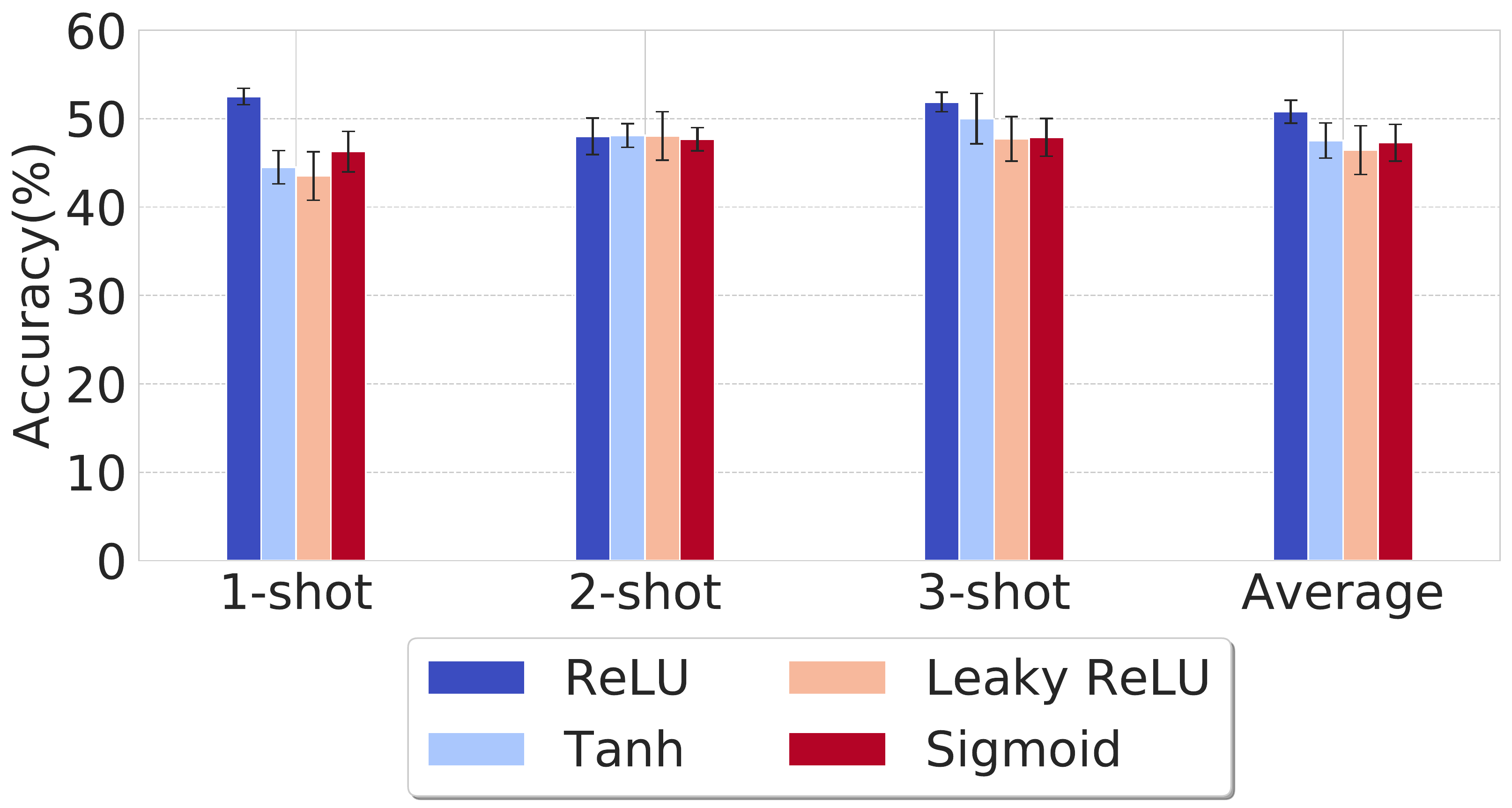}
	    \vspace{-1ex}		
		\caption{Activation function selection.}
		\label{fig:activation}
	\end{figure}

    \subsection{Hyperparameter Searching} 
    We first conduct hyperparameter searching of dual-path base network, and thus we could employ the final hyperparameters in the following experiments for the fairness comparison. To be specific, we explore the following aspects: selection of CNN backbone, hidden dimension size, as well as activation function selection. In this section, we explore the aspects on the dual-path base network on the Wi-Fi dataset with 100 environments.
	
	\subsubsection{Backbone Selection} \label{sssec:backbone}
    Since our base network leverages CNN backbone $f_b$ for extracting spatial features, the selection of CNN-backbone is essential for building a good base network. Therefore, we evaluate multiple state-of-the-art CNN backbones with various model complexity implanted into our base network. As depicts in Figure~\ref{fig:backbone}, these include two shallow networks: 5 layers CNN (cnn5) (with 3 convolutional layers and 2 fully connected layers) and 8 layers AlexNet (alexnet8)~\cite{krizhevsky2012imagenet}, two medium deep networks: 18 layers ResNet (resnet18)~\cite{he2016deep} and 19 layers VGG (vgg19)~\cite{simonyan2015very}, as well as two deep networks: 44 layers ShuffleNet (shufflenet44)~\cite{Zhang20186848} and 152 layers ResNet (resnet152)~\cite{he2016deep}. We observe that our choice resnet18 achieves the best overall accuracy compared with other backbones, particularly for one-shot HAR. We further study the model overhead (complexity, inference time, and memory usage) of resnet. As the complexity of resnet18 takes up around 85\% of the whole system, its overhead can be roughly regarded as the system overhead. According to a recent benchmark~\cite{bianco2018benchmark}, resnet18, with a low model complexity and memory usage, achieves a low inference time, leading to an excellent real-time performance. Moreover, compared with other light-weight model such as shufflenet44, resnet18 achieves a comparable performance when it is deployed to commercial edge devices. Therefore, we select pre-trained ResNet 18 layers network as the CNN backbone for \systemname. 

	\subsubsection{Hidden Dimension Size \& Activation Function Selection}
	As described in Section~\ref{sec:har}, we use a hidden variable $\alpha$ to compose our network. Therefore, we aim to search a good dimension size. We separately explore three settings $\alpha = 128$, $\alpha = 256$, and $\alpha = 512$, and report the accuracy in Figure~\ref{fig:hidden_dim}. \textcolor{black}{Although $\alpha = 128$ is compared well with $\alpha = 512$ in average, $\alpha = 512$ outperforms $\alpha = 128$ on one-shot HAR. In this paper, one-shot HAR takes precedence, and thus we set $\alpha$ to 512.} Afterwards, we also search the best activation function in our network. We explore ReLU, Leaky ReLU, Sigmoid, and Tanh. Figure~\ref{fig:activation} plots the performance. We observe that ReLU outperforms other function and we set all the activation function to ReLU. Therefore, we will employ pre-trained ResNet with 18 layers as our CNN backbone, set hidden dimension to 512 and select ReLU as activation function in the following experiments.

    \subsection{Superiority of Proposed Network}
    In this section, we will investigate the superiority of dual-path base network and {\systemname} over baselines on various RF sensing datasets. Specifically, we set following parameters: $M'$ to 12, batch size to 3, learning rate to 0.001, the number of epochs to 20 and we employ Adam optimizer to train networks. 

    \subsubsection{Superiority of Dual-path base Network over Baselines}
\begin{figure}[t]
		\centering
	    \includegraphics[width = 0.43\textwidth]{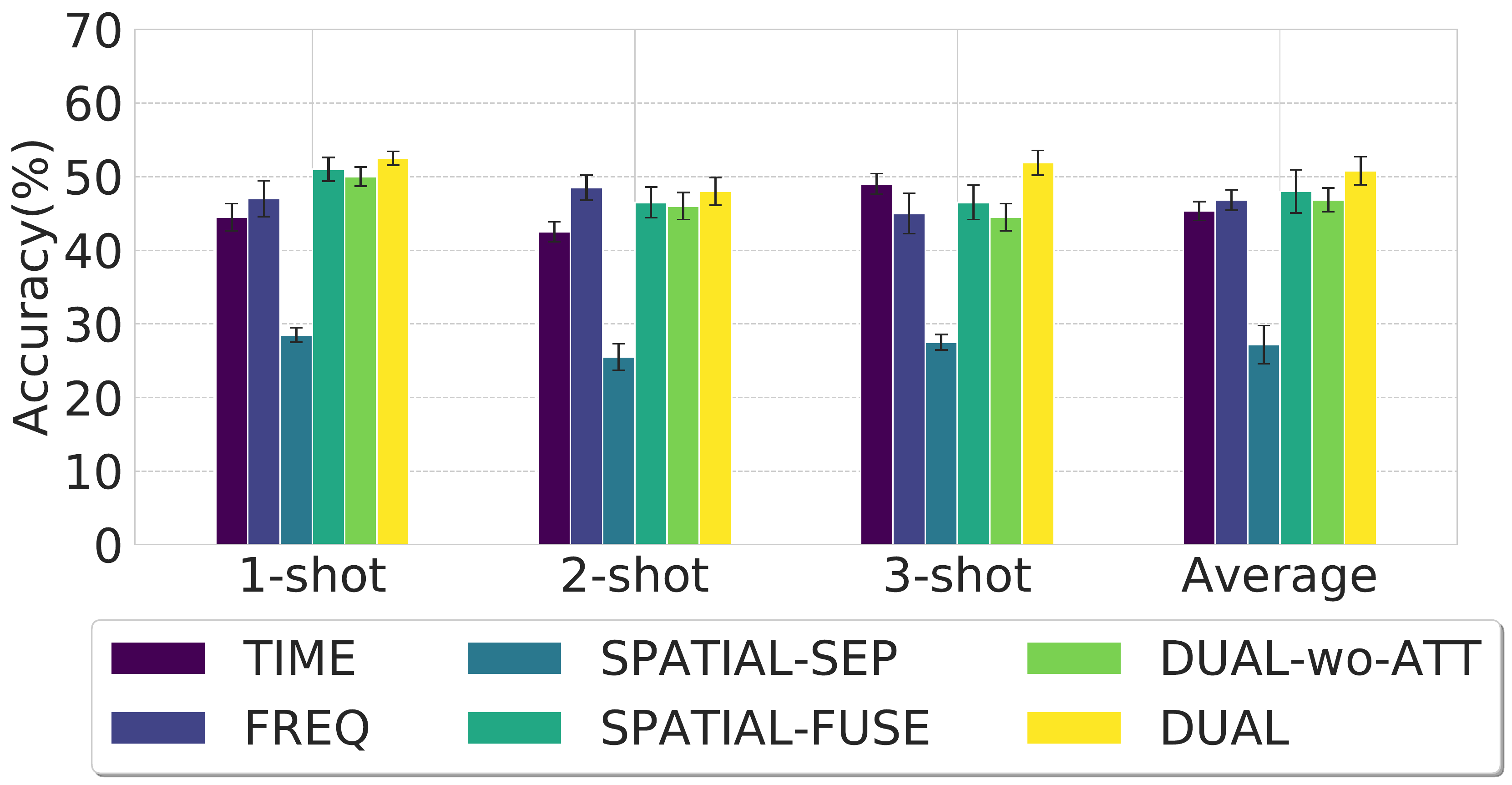}
	    \vspace{-1ex}
		\caption{Overall comparison results for dual-path base network evaluation on Wi-Fi dataset.}
		\label{fig:wifi-har}
	\end{figure}

\begin{figure}[t]
		\centering
	    \includegraphics[width = 0.43\textwidth]{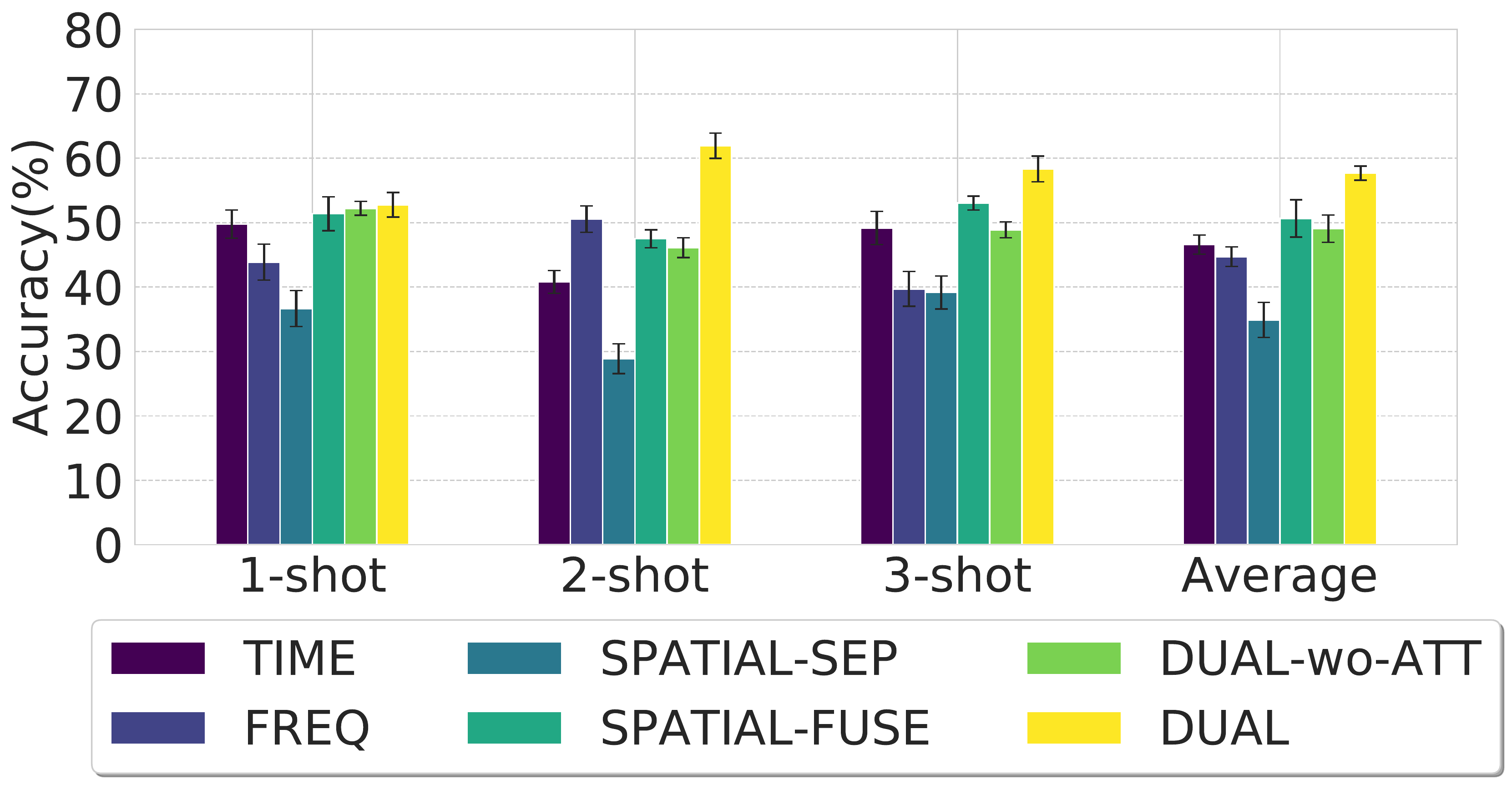}
	    \vspace{-1ex}
		\caption{Overall comparison results for dual-path base network evaluation on FMCW dataset.}
		\label{fig:fmcw-har}
	\end{figure}

\begin{figure}[t]
		\centering
	    \includegraphics[width = 0.43\textwidth]{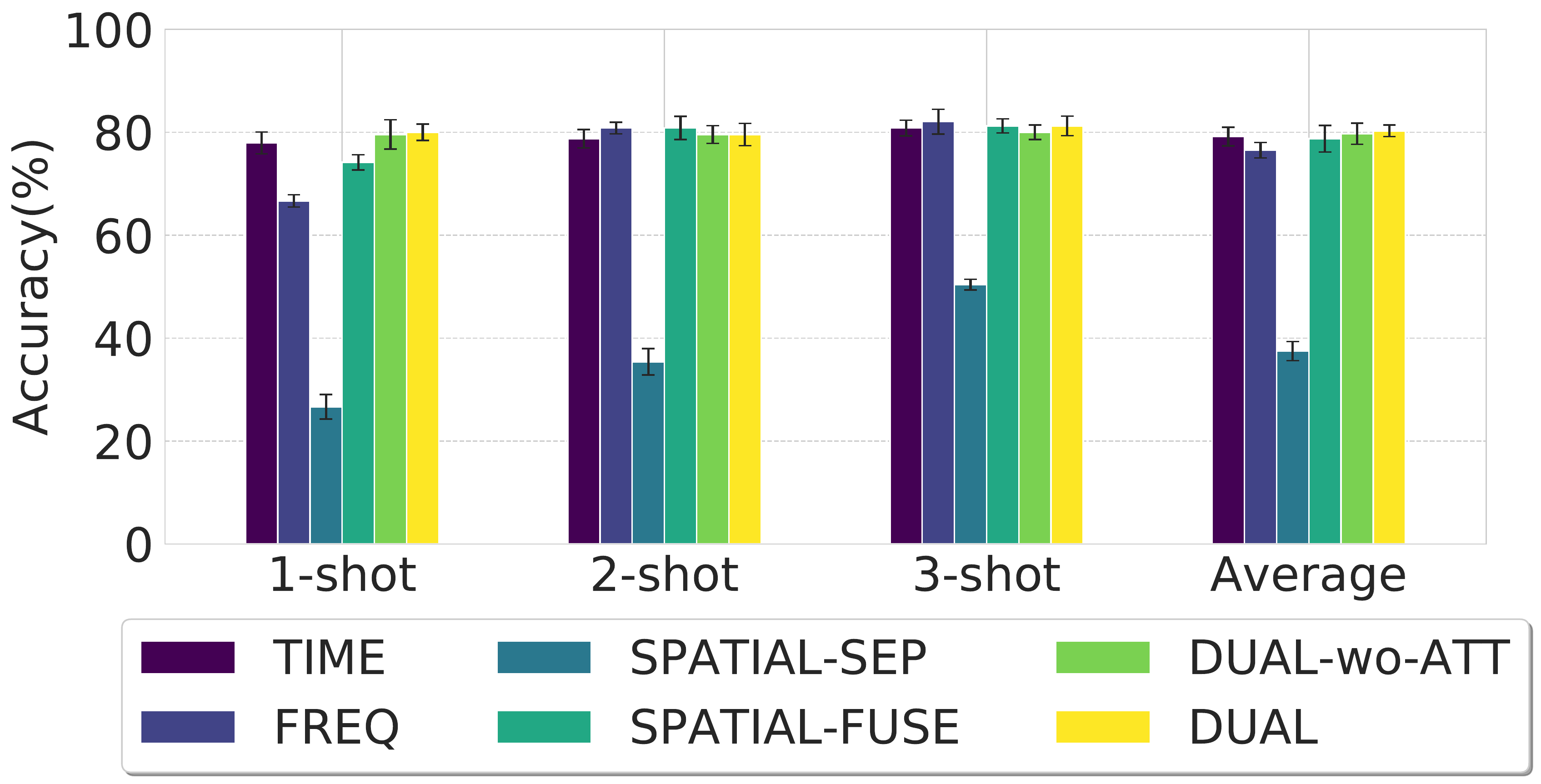}
	    \vspace{-1ex}
		\caption{Overall comparison results for dual-path base network evaluation on IR dataset.}
		\label{fig:uwb-har}
	\end{figure}

    Figure~\ref{fig:wifi-har}, Figure~\ref{fig:fmcw-har}, and Figure~\ref{fig:uwb-har} separately report the accuracy of dual-path base network and the baselines on Wi-Fi, FMCW, and IR datasets. In all the cases, our base network outperforms all the baselines. It shows the effectiveness of both spatial and temporal features extracted by the dual-path base network for the purpose of HAR. To be specific, the performance of SPATIAL-FUSE outperforms TIME and FREQ remarkably. It validates our motivation that the utilization of RF signal matrix in dual-path (both time and frequency domain) is beneficial for HAR. \textcolor{black}{In our experiments, we find that SPATIAL-FUSE is far better than SPATIAL-SEP when accuracy and time complexity are taken into account}. One possible reason is that using the CNN backbone twice doubles the number of parameters in the network, and makes SPATIAL-SEP relatively harder to train and converge.  
	%
	\begin{figure}[t]
		\vspace{-2ex}
		\centering
		\subfloat[Wiping (frequency)]{
		    \begin{minipage}[b]{0.48\linewidth}
		        \centering
			    \includegraphics[width = .96\textwidth]{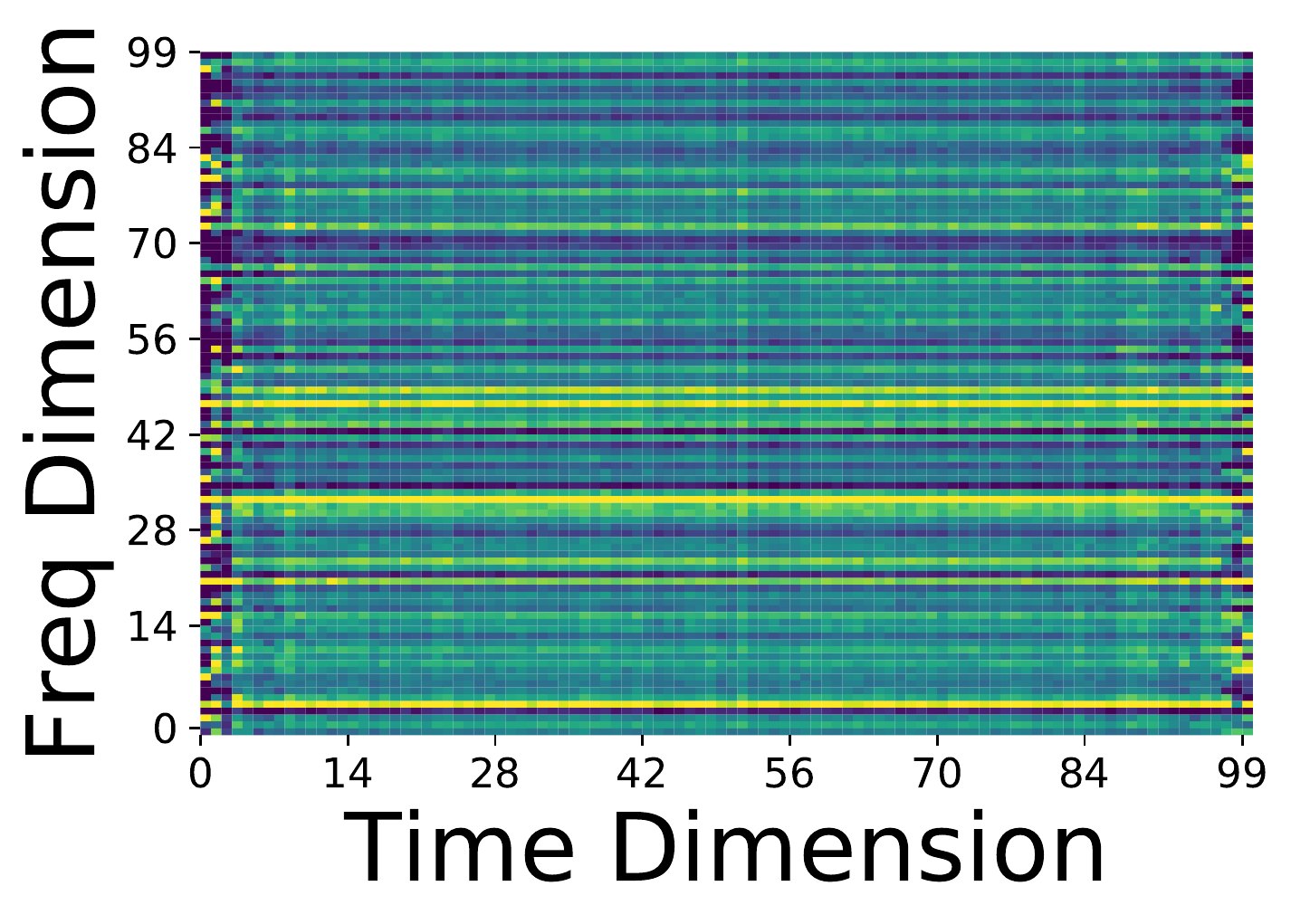}
			    \vspace{-.5ex}
			\end{minipage}
		}%
		\subfloat[Walking (frequency)]{
		    \begin{minipage}[b]{0.48\linewidth}
		        \centering
			    \includegraphics[width = .96\textwidth]{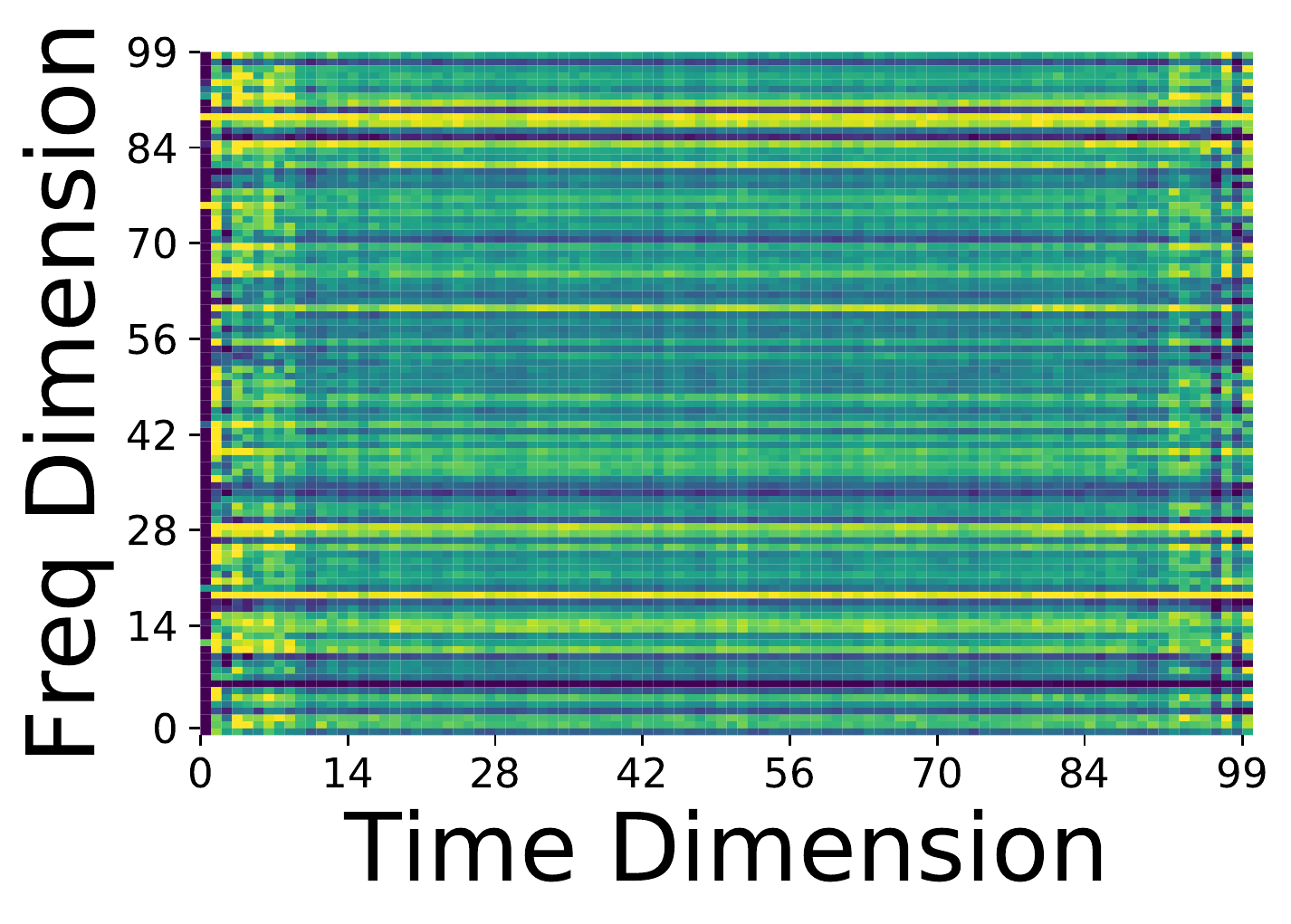}
			    \vspace{-.5ex}
			\end{minipage}
		}
	    \vspace{-1ex}
		\caption{FMCW attention map $\mathcal{A}$ of two different activities: walking and sitting down.}
		\label{fig:attention}
	    \vspace{-1ex}
	\end{figure}
	\begin{figure}[t]
		\centering
	    \includegraphics[width = 0.43\textwidth]{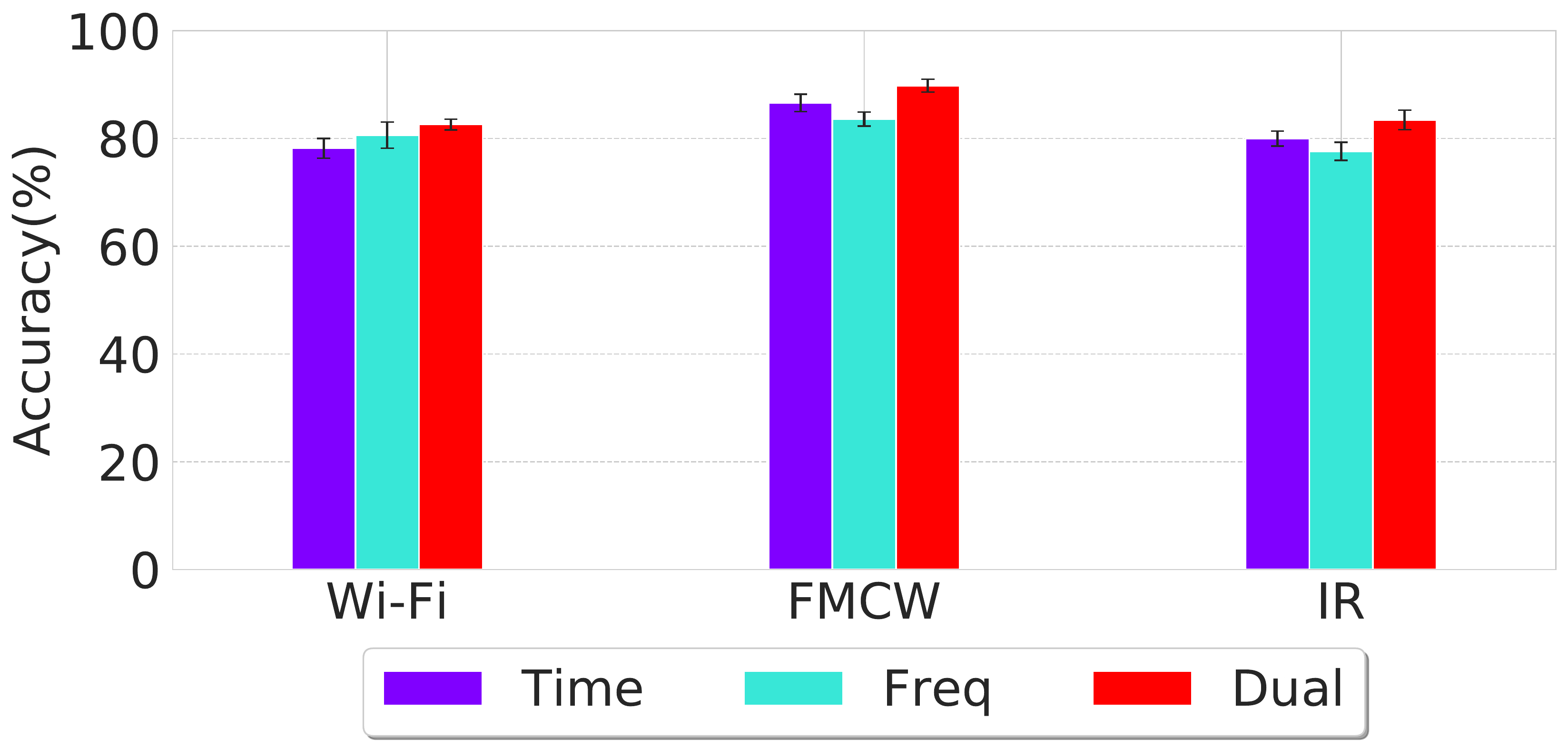}
	    \vspace{-1ex}
		\caption{Overall comparison results for dual-path base network evaluation on Wi-Fi, FMCW, IR datasets in terms of general classification.}
		\label{fig:noshots}
	    \vspace{-2ex}
	\end{figure}

    
    In most cases, we observe that the performance of DUAL-wo-ATT is still unacceptable in spite of already involving temporal features. By contrast, the outstanding performance of dual-path base network demonstrates the efficacy of attended temporal features. We further plot attention map $\mathcal{A}$ of two different activities on FMCW datasets in Figure~\ref{fig:attention}. This showcases that the attention mechanism used in the temporal module is able to discover a rich joint temporal representations across two domains, thus improving the performance of HAR. In general, the accuracy of our dual-path base network outperforms DUAL-wo-ATT by around 4.2\%. 
    
    \vspace{1ex}
	\noindent\textbf{Remark}: 
	One may intuitively expect a monotonic increase in accuracy given more shots. However, as the base network is designed for RF-HAR in better exploiting the input of high-dimensional RF signal matrices, its performance is optimized for general classification to fit into the overall meta learning framework. Therefore, the fluctuation in accuracy across a few shots is reasonable and can be expected. We further demonstrate the efficacy of dual-path base network for general classification. As shown in Figure~\ref{fig:noshots}, dual-path base network performs still outstanding consistently in general classification on all RF datasets. Figure~\ref{fig:noshots}  (albeit not a fair comparison due to the separated datasets for different RF sensing techniques) also sheds a light on the intrinsic differences among the three RF signals. Wi-Fi with narrowband cannot provide a high resolution in time domain, but Doppler shift can still be retained in frequency domain. Therefore, the overall performance of Wi-Fi is worse than the others, but the accuracy in frequency domain is better than that in time domain. Resorting to the power of wideband, FMCW and IR both achieve better performance due to a higher time resolution, yet FMCW slightly outperforms IR thanks to a higher tx power. 

    \subsubsection{Superiority of {\systemname} over Baselines}
\begin{figure}[t]
		\centering
	    \includegraphics[width = 0.43\textwidth]{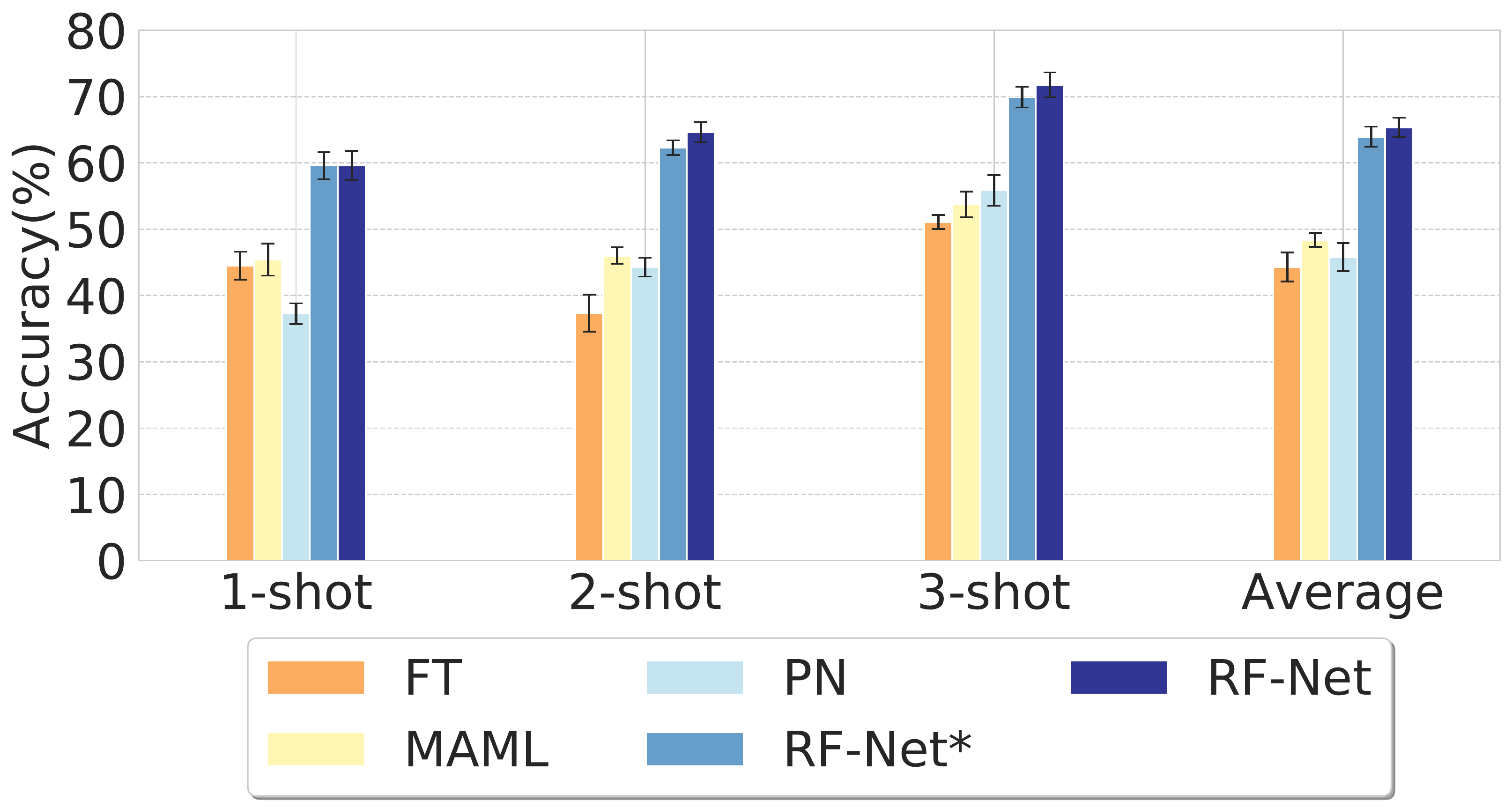}
	    \vspace{-1ex}
		\caption{Overall comparison results for \systemname\ evaluation on Wi-Fi dataset.}
		\label{fig:wifi-rf}
	    \vspace{-1ex}
	\end{figure}

\begin{figure}[!t]
		\centering
	    \includegraphics[width = 0.43\textwidth]{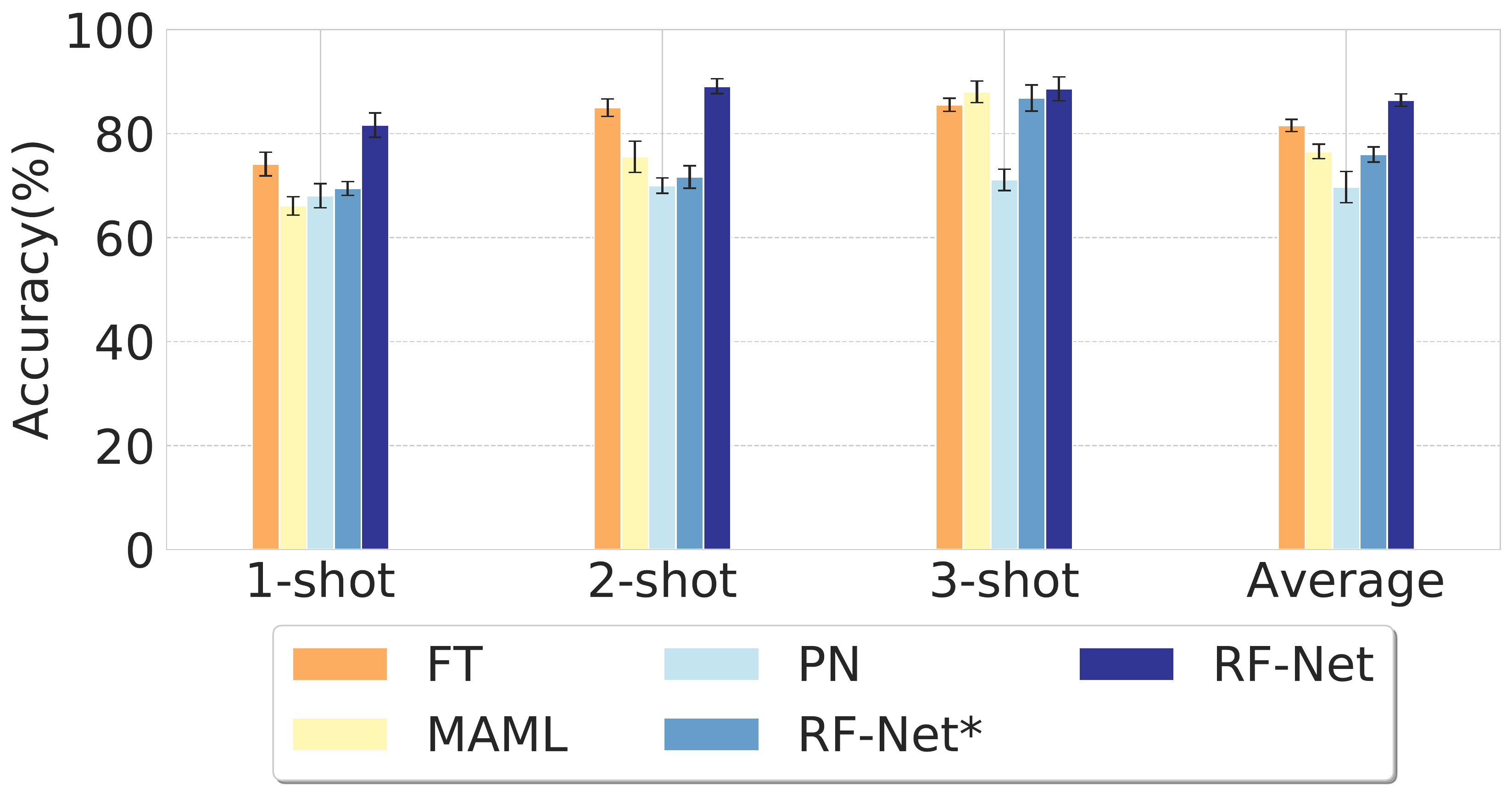}
	    \vspace{-1ex}
		\caption{Overall comparison results for \systemname\ evaluation on FMCW dataset.}
		\label{fig:fmcw-rf}
	    \vspace{-1ex}
	\end{figure}

\begin{figure}[!t]
		\centering
	    \includegraphics[width = 0.43\textwidth]{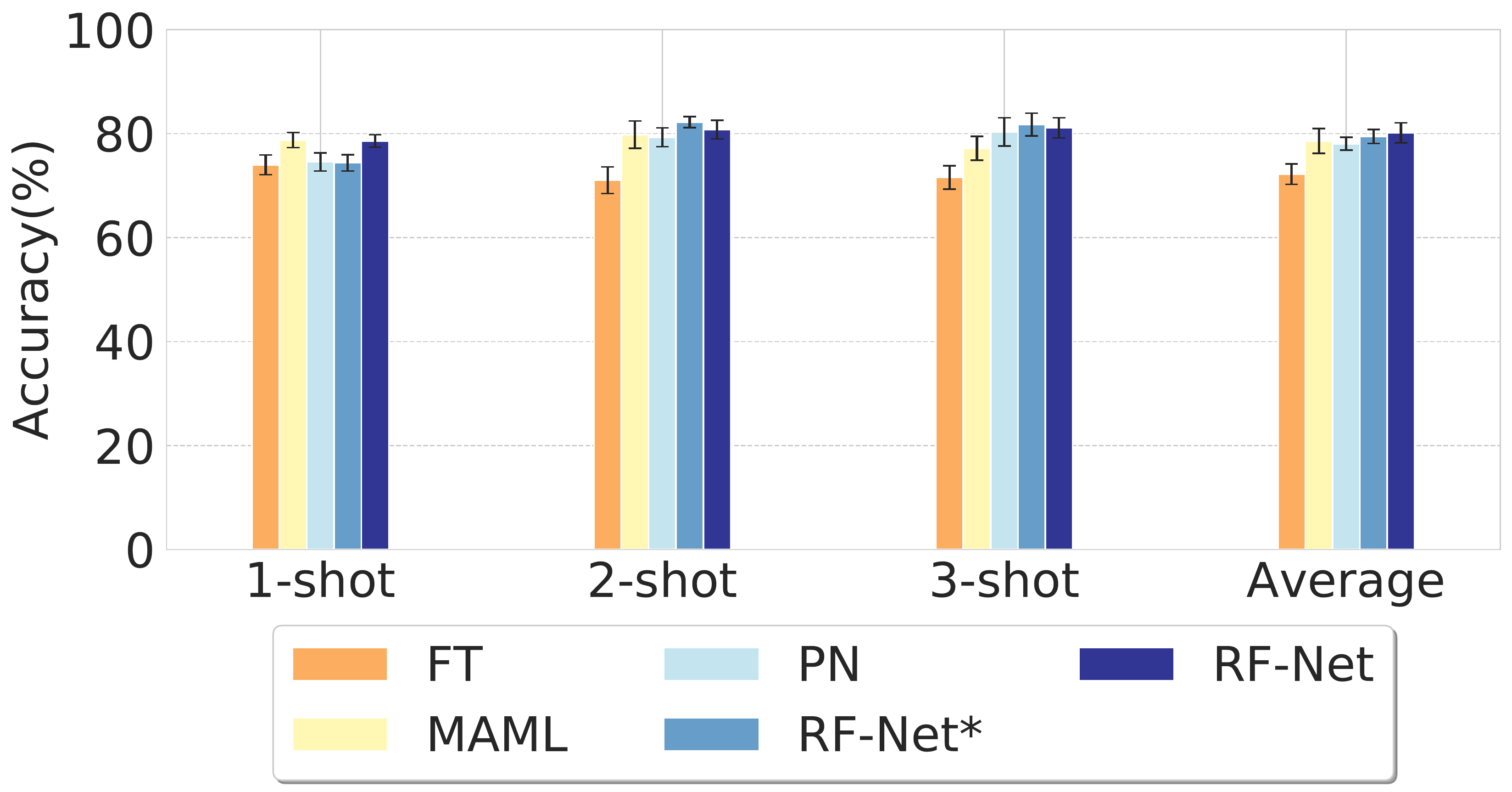}
	    \vspace{-1ex}
		\caption{Overall comparison results for \systemname\  evaluation on IR dataset.}
		\label{fig:uwb-rf}
	    \vspace{-1ex}
	\end{figure}

    The comparison results for {\systemname} on Wi-Fi, FMCW, and IR are listed in Figures~\ref{fig:wifi-rf},~\ref{fig:fmcw-rf}, and~\ref{fig:uwb-rf}. We observe that our proposed {\systemname} network has proved the superiority over baselines on all three RF sensing techniques. In particular, compared {\systemname} with {\systemname}*, overall performance is enhanced by RF metric module and relatively steady. It highlights the efficacy of RF metric module. This performance improvement gives credit to the trainable linear mapping of RF metric module. As demonstrated in Section~\ref{sssec:rf_metric}, its convex loss function should allow the RF metric module to learn an optimal 
    metric for \systemname. One may argue that all three RF techniques achieve a rather low accuracy, but this is the best one may obtain so far when adapting to new environments with very few labelled observations; in fact, our performance is already better than that for image recognition~\cite{chen2018closer}, possibly thanks to the depth information obtained by RF sensing. 
							
\begin{figure}[!t]
		\centering
	    \includegraphics[width = 0.432\textwidth]{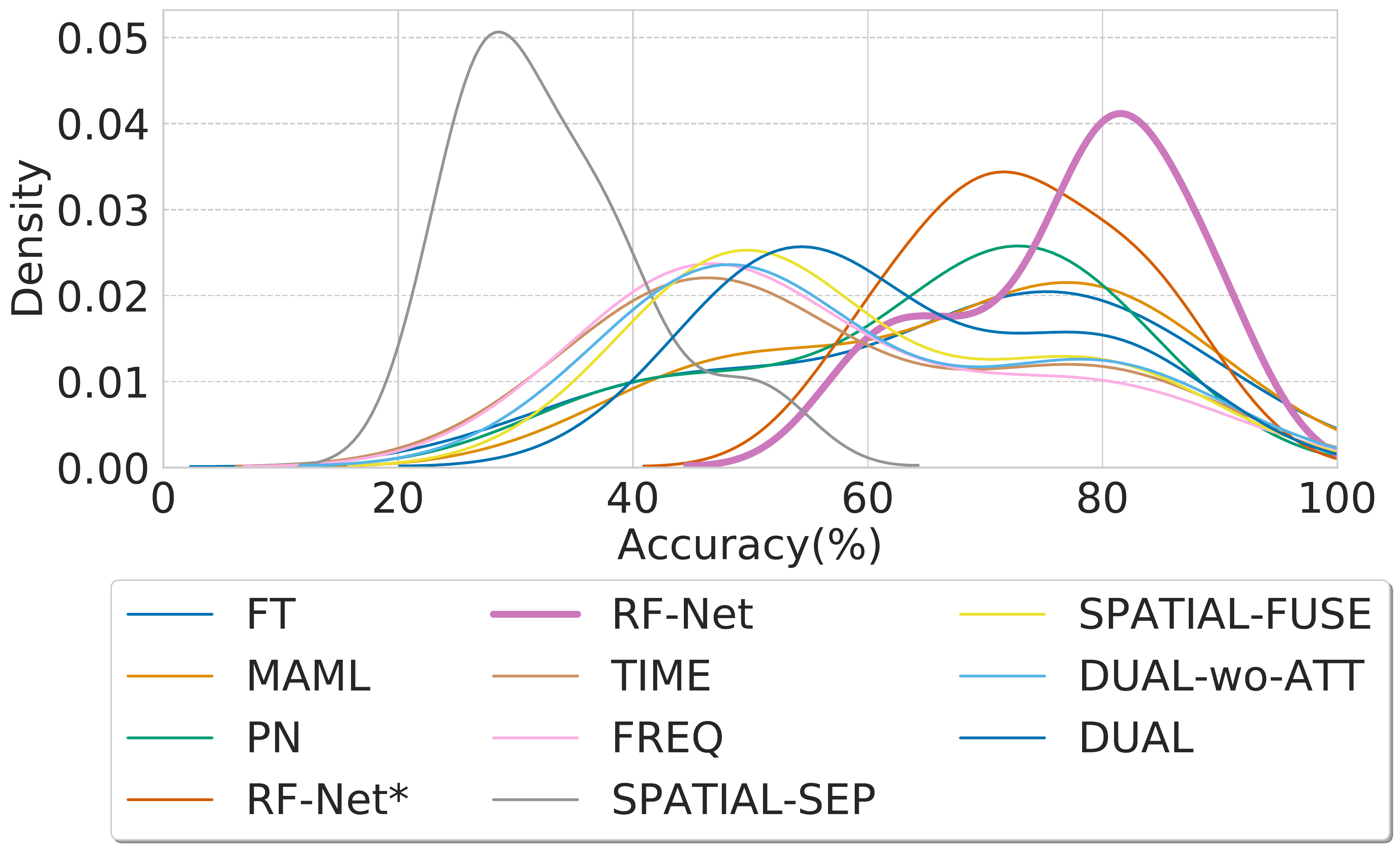}
	    \vspace{-1ex}
		\caption{Accuracy density plot of all evaluated networks.}
		\label{fig:density}
	\end{figure}

    In Figure~\ref{fig:density}, we further plot the accuracy density plot of {\systemname} along with the dual-path base network and all the baselines. The best performance is highlighted in bold. The density plot depicts the accuracy distribution on all three RF datasets with $N_s = \{1,2,3\}$. Intuitively, dual-path base network notably outperforms its baselines. However, it is still far below the expectation. Compared with dual-path base network, the density plot of {\systemname} along with its meta-learning baselines and fine-tuning baselines exhibit some notable right skew. It indicates that {\systemname} and its baselines as expected entail significantly performance improvement of the implanted base network. Nevertheless, {\systemname} boosts the performance maximally among them: it has greatly helped the base network adapting to new environment with even a single observation. 
    
 	

	\begin{figure}[t]
		        \centering
			    \includegraphics[width = .48\textwidth]{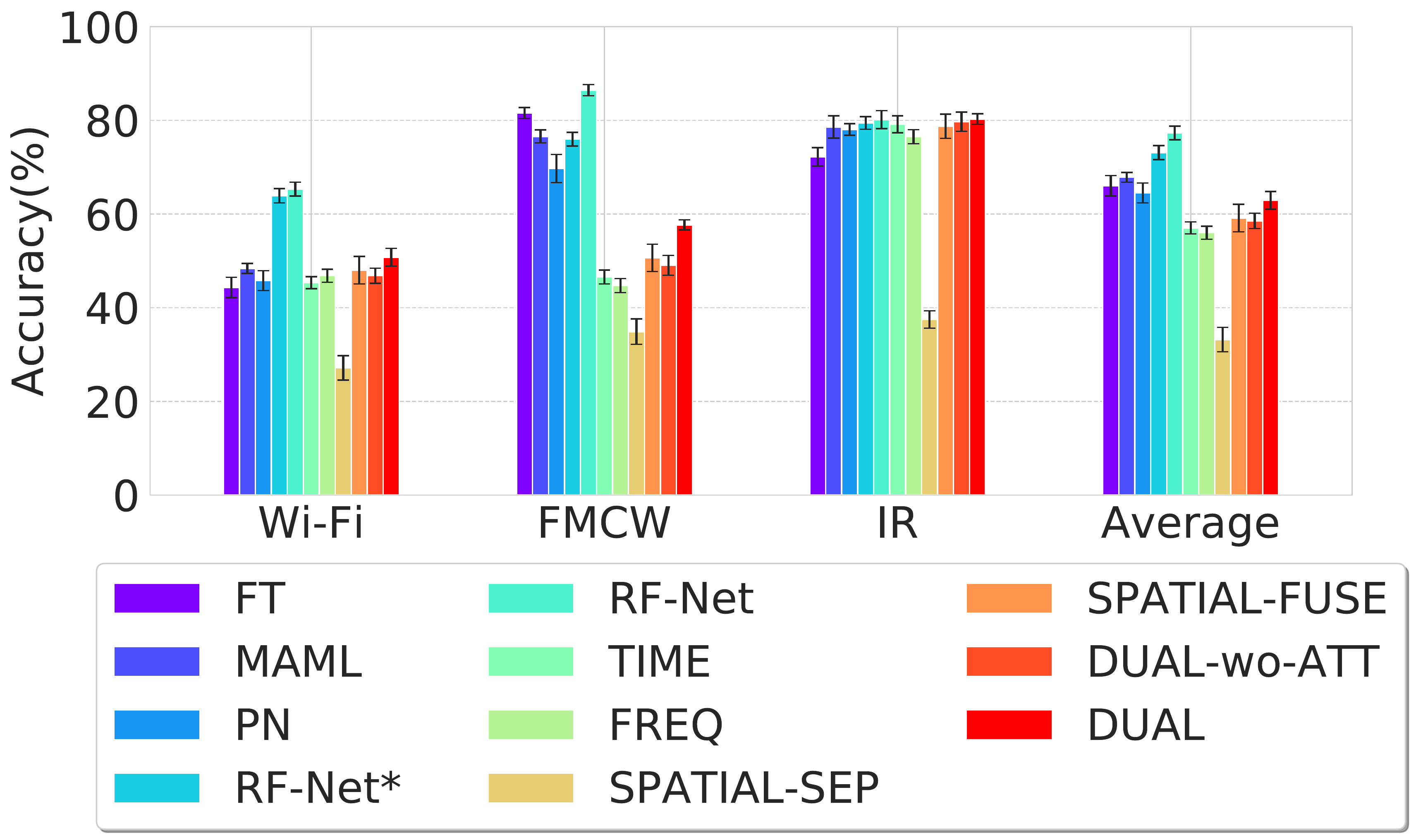}
	    \vspace{-4ex}		
		\caption{Performance on RF sensing types.}
		\label{fig:acc_sense}
	\end{figure}	
		
	\begin{figure}[b]
		\centering
	    \includegraphics[width = 0.49\textwidth]{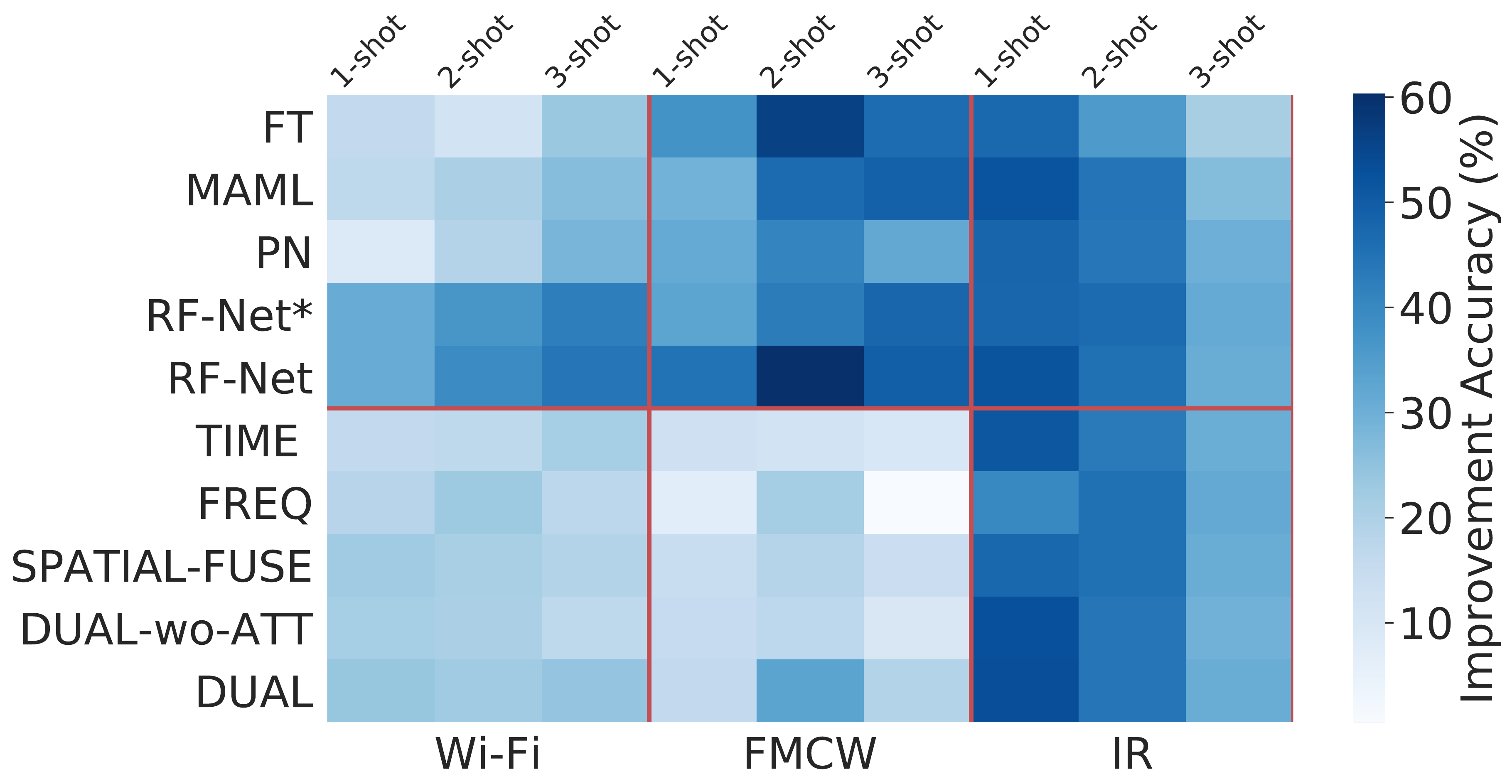}
	    \vspace{-4ex}
		\caption{Performance heatmap: accuracy improvement from SPATIAL-SEP, while a darker color denotes a higher accuracy improvement.}
		\label{fig:perf_heatmap}
	\end{figure}

    \subsubsection{Wi-Fi vs. IR vs. FMCW}
To further evaluate the performance on each RF sensing dataset, we first plot the performance of the proposed networks and their baselines on Wi-Fi, IR, and FMCW in Figure~\ref{fig:acc_sense}. Meanwhile, we further generate a more representative performance heatmap in Figure~\ref{fig:perf_heatmap} to help visualizing varying degrees of improved performance by {\systemname} in each RF sensing dataset. From these two figures, we observe that \systemname\ \textcolor{black}{achieves the highest overall performance in all datasets, and the performance of all concerned networks differ significantly across different RF sensing datasets}. One curious observation on the evaluation over IR dataset is that all networks seem to achieve similar performances with and without meta-learning.
The reason is that the intrinsic advantage of IR being not sensitive to environment impact (thus allowing the environment-free features to be captured via base network), yet RF-Net still performs the best among all networks.
%
To summarize, based on the average accuracy shown in Figures~\ref{fig:acc_sense} and \ref{fig:perf_heatmap}, our proposed RF-Net are demonstrated to offer the most robust generalization that may fit all data intrinsic properties of all RF signals.

	\begin{figure}[t]
		\centering
	    \includegraphics[width = 0.48\textwidth]{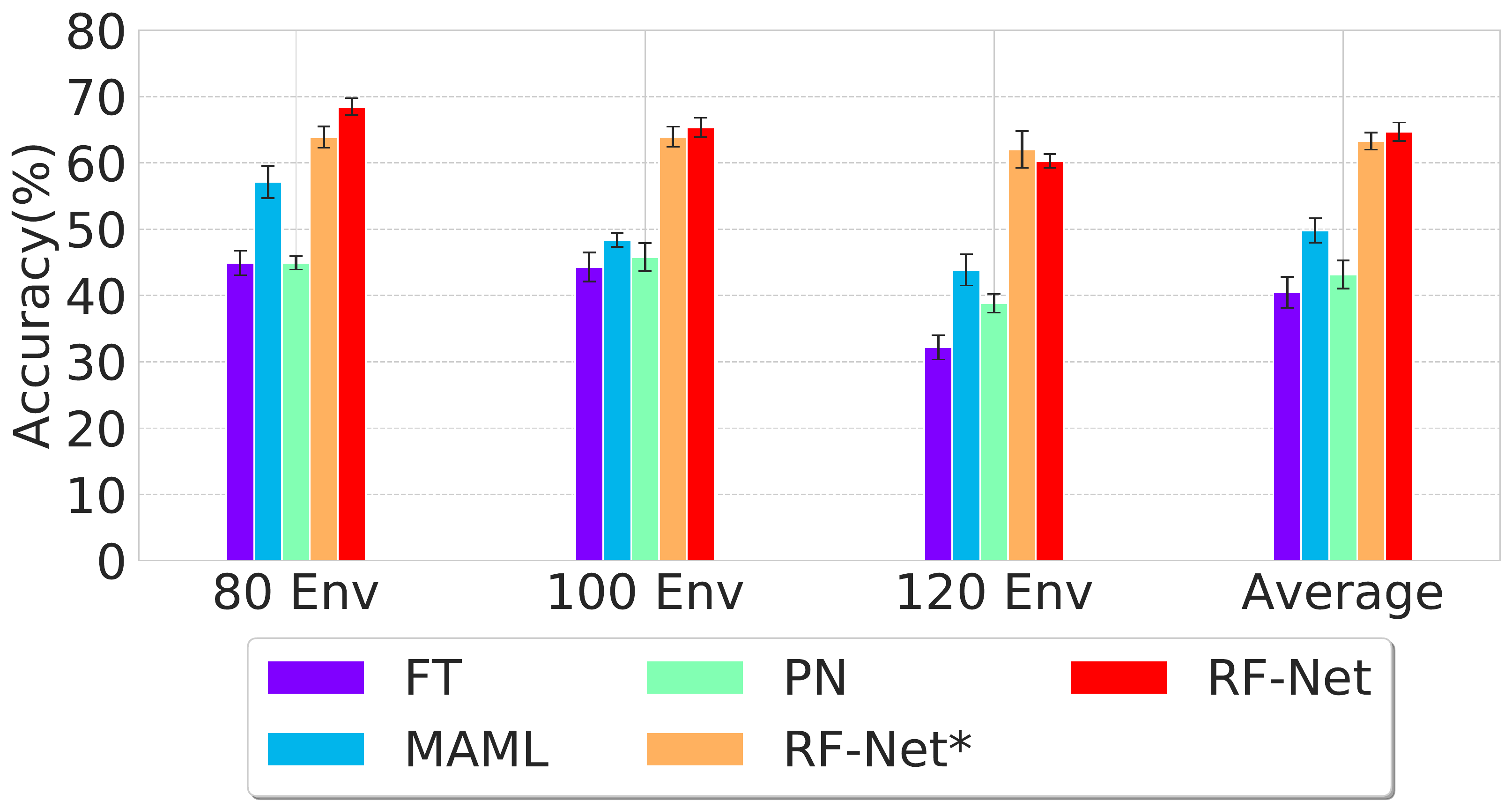}
	    \vspace{-4ex}
		\caption{Impact of number of environments.}
		\label{fig:num_env}
	\end{figure}
    
    \subsubsection{Impact of Number of Environments}
    We examine networks from the perspective of environmental diversity as well. The motivation is to explore how many environments could be handled for each network. We report average accuracy of {\systemname} and its baselines on Wi-Fi dataset with 80, 100 and 120 environments in Figure~\ref{fig:num_env}. With the number of environments increasing, we observe that the performance of both fine-tuning and meta-learning baselines (MAML and PN) have clearly dropped. Our {\systemname} show more steady trends. \textcolor{black}{This performance indicates that {\systemname} can deal with more environments without sacrificing performance severely, i.e., it is a more robust approach.}
    

	\section{Conclusion } \label{sec:con}
Based on the study of representative RF sensing techniques along with major meta-learning approaches, we have proposed {\systemname}, a meta-learning based neural network for one-shot RF-based HAR; it contributes to the capability of being fast adapted to new environments with a single observation. {\systemname} consists of a meta-learning framework that involves a parametric RF-specific module for training a powerful distance metric, and a dual-path base network that fully exploits the high-dimensional features contained in the RF signal matrix. We have conducted extensive experiments on all three RF sensing techniques. These experiments have demonstrated the efficacy of our proposed {\systemname} and dual-path base network.

As a potential future direction, we are looking into extending our meta-learning frameworks to various sensing applications that can be heavily affected by environment changes. These applications include, among others, radio frequency sensing for, e.g., vibration detection and indoor localization~\cite{UWHear,luo2014ilocscan,Zhe2017awl}, acoustic sensing for similar purposes~\cite{FollowMeDrone,SST,AcuTe}, and visible light sensing for, e.g., information decoding~\cite{CeilingCast,ReflexCode} and occupancy inference~\cite{CeilingSee,YANG201835}. As these applications may involve rather different sensing methodologies and modalities, both the base network and the meta-learning framework will need to be substantially revamped.
	
	\section{Acknowledgements}
We are grateful to anonymous shepherd and reviewers for their valuable comments. This research is supported in part by AcRF Tier 2 Grant MOE2016-T2-2-022 and AcRF Tier 1 Grant RG17/19. 

\newpage
	
	\bibliographystyle{acm}
	\bibliography{acmart}
	
\end{document}